\documentclass[11pt,a4paper,english]{article}
\usepackage{babel,cite}

\makeatletter

\providecommand{\LyX}{L\kern-.1667em\lower.25em\hbox{Y}\kern-.125emX\@}

\usepackage{amssymb,amsthm}

\newcommand{\be}{\begin{equation}}
\newcommand{\ee}{\end{equation}}
\newcommand{\eq}[1]{(\ref{#1})}
\def\beq{\be}
\def\eeq{\ee}

\def\nn{\nonumber}
\def\bea{\begin{eqnarray}}
\def\eea{\end{eqnarray}}
\newcommand{\barr}{\begin{array}}
\newcommand{\earr}{\end{array}}
\def\obar{\overline}
\def\one{\mbox{1 \kern-.59em {\rm l}}}

\def\a{\alpha}          
\def\b{\beta}           
  \def\C{\Gamma}  
\def\d{\delta}    
\def\e{\epsilon}                \def\vare{\varepsilon}
        
\def\g{\gamma}   \def\G{\Gamma}

 \def\la{\lambda}

\def\cA{\mathcal{A}}
\def\C{{\mathbb C}}
\def\cN{\mathcal{N}}

\def\R{{\mathbb R}}
\def\C{{\mathbb C}}

\def\Z{{\mathbb Z}}
\def\one{\mbox{1 \kern-.59em {\rm l}}}
\newcommand{\torus}{{\mathbb T}}

\def\bit{\begin{itemize}}
\def\eit{\end{itemize}}

\def\tens{\otimes}

\def\reps{representations }

\def\pp#1{\partial_#1}
\def\mc#1{\mathcal{#1}}

\def\dd{\partial}

\makeatother
\begin{document}

\renewcommand{\title}[1]{\vspace{10mm}\noindent{\Large{\bf #1}}\vspace{8mm}}
\newcommand{\authors}[1]{\noindent{\large #1}\vspace{5mm}}
\newcommand{\address}[1]{{\itshape #1\vspace{2mm}}}

\begin{titlepage}

\begin{flushright}
hep-th/0503041\\
LMU-ASC 17/05   \\
MPP-2005-14
\end{flushright}

\begin{center}
  
\title{ \LARGE Gauge Theory on Fuzzy $S^{2}\times S^{2}$ \\ and \\[4pt]
  Regularization %of Gauge Theory 
 on Noncommutative $\R^4$ \vspace{1cm} }

\authors{Wolfgang {\sc Behr$^{a,b}$},\  Frank {\sc Meyer$^{a,b}$} \  and Harold {\sc Steinacker}$^a$ \vspace{0.8cm}}

\address{$^{a}$\, Arnold Sommerfeld Center, Department f\"ur Physik\\
\it Ludwig-Maximilians-Universit\"at M\"unchen, Theresienstra{\ss}e 37\\
\it D-80333 M\"unchen, Germany
}

\address{$^{b}$\,Max--Planck--Institut f\"ur Physik\\ \it (Werner-Heisenberg Institut) \\
\it F\"ohringer Ring 6, D-80805 M\"unchen, Germany
\\}

\address{{\it E--mail:} {\tt behr,meyerf,hsteinac@theorie.physik.uni-muenchen.de}}

\vskip 2cm

\textbf{Abstract}

\vskip 3mm

\begin{minipage}{14cm}%

We define $U(n)$ gauge theory on fuzzy $S^{2}_N\times S^{2}_N$
as a multi-matrix model, which reduces to ordinary Yang-Mills theory
on $S^{2}\times S^{2}$ in the commutative limit $N \to\infty$.
The model can be used as a  regularization of gauge theory on
noncommutative 
$\R^4_\theta$ in a particular scaling limit, which is studied in
detail. We also find topologically non-trivial $U(1)$ solutions, 
which reduce to the known ``fluxon'' solutions in the
limit of $\R^4_\theta$, reproducing their full moduli space. 
Other solutions which can be interpreted as
2-dimensional branes are also found. 
The quantization of the model is defined non-perturbatively
in terms of a path integral which is finite.
A gauge-fixed BRST-invariant action is given as well. 
Fermions in the fundamental representation of the gauge group are included
using a formulation based on $SO(6)$, by
defining a fuzzy Dirac operator which reduces to the
standard Dirac operator on $S^{2}\times S^{2}$ in the
commutative limit. The chirality operator 
and Weyl spinors are also introduced.

\end{minipage}

\end{center}

%\vspace{1cm}
%{Keywords:} Nonperturbative Effects, Matrix Models, Non-Commutative Geometry

\end{titlepage}

 \tableofcontents

\section{Introduction}

Gauge theories on noncommutative spaces have received much attention
in recent years. One of the reasons is the natural
realization of such theories in the framework of string theory and
$D$-branes \cite{Seiberg:1999vs}, however they deserve interest also
in their own right; see \cite{Douglas:2001ba,Szabo:2001kg} for some reviews.
One of the most remarkable new features of  noncommutative gauge 
theories is the fact that they can be defined in terms of multi-matrix
models, which means that the action involves only products of  
``covariant coordinates'' $X_i = x_i + A_i$, with gauge transformations
acting as $X_i \to U X_i U^{-1}$. 
In particular for certain quantized compact spaces 
such as fuzzy spheres and tori,
these $X_i$ are finite-dimensional Hermitian matrices of size $N$. 
Nevertheless, the conventional gauge theory is correctly reproduced in the 
limit $N \to \infty$.
This leads to a natural quantization prescription by simply
integrating over these matrices.
For the much-studied case of the quantum plane 
$\R^d_\theta$, the matrices $X_i$ are infinite-dimensional, and
the precise definition 
of the models is quite non-trivial. This is particularly obvious
by noting that the naive action for gauge theory on $\R^d_\theta$
contains sectors with any rank of the gauge group $U(n)$ \cite{Gross:2000ss}.
To have a well-defined theory and quantization prescription,
a regularization of gauge
theory on $\R^d_\theta$ based on the finite
compact case is therefore very desirable. Furthermore, the formulation
as multi-matrix model leads to the hope that non-trivial results may
be obtained using the sophisticated techniques from random matrix theory.
We introduce in this paper such a matrix model for fuzzy $S^2 \times
S^2$, and study its relationship with $\R^4_\theta$.

In the 2-dimensional case, this matrix-model approach
to gauge theory has been studied in
considerable detail for the fuzzy sphere $S^2_N$ 
\cite{Madore:1992,Carow-Watamura:1998jn,Steinacker:2003sd,Presnajder:2003ak,Imai:2003vr,Castro-Villarreal:2004vh} and the 
noncommutative torus $\torus^2_\theta$ 
\cite{Ambjorn:1999ts,Paniak:2002fi,Paniak:2003xm,Griguolo:2003kq}, both on the
classical and quantized level.
It is well-known that $\R^2_\theta$ can be obtained
as scaling limits of these spaces $S^2_N$ and $\torus^2_N$ at least
locally, which suggests a correspondence also for the 
gauge theories. This correspondence of gauge theories
has been studied in great detail for the case of 
$\torus^2_\theta\to \R^2_\theta$  
\cite{Paniak:2002fi,Griguolo:2001ce,Griguolo:2004jp} on the
quantized level, exhibiting
the role of certain instanton contributions. 
A matching of gauge theory on the classical level
can also be seen for $S^2_N \to \R^2_\theta$
\cite{Syko:unpublished,Ydri:2004im}, 
which is implicitly 
contained in Section \ref{sec:scaling} of the present paper. 

In 4 dimensions, 
the quantization of gauge theory is more difficult, and a regularization using 
finite-dimensional matrix models is particularly
important. The most obvious 4-dimensional spaces 
suitable for this purpose are 
$\torus^4, S^2 \times S^2$ and $\C P^2$. 
On fuzzy $\C P^2_N$ 
\cite{Grosse:1999ci,Alexanian:2001qj,Carow-Watamura:2004ct}, such
a formulation of gauge theory was given in 
\cite{Grosse:2004wm}. This
can indeed be used to obtain $\R^4_\theta$ for the case of
$U(2)$ -invariant $\theta_{ij}$. 
The case of $\R^2 \times S^2_N$ as regularization of $\R^4_\theta$
with degenerate $\theta_{ij}$ 
was considered in \cite{Ydri:2004im,Ydri:2004vq}, exhibiting a relation with a
conventional non-linear sigma model.
A formulation of lattice gauge theory 
for even-dimensional tori has  been discussed in
\cite{Ambjorn:2000nb,Ambjorn:2000cs,Griguolo:2003kq}.
Related ``fuzzy'' solutions of the string-theoretical matrix 
models \cite{Ishibashi:1996xs} were studied e.g. 
in \cite{Iso:2001mg,Kitazawa:2002xj}, see also \cite{Kimura:2002nq}. 

In the present paper we give a
definition of $U(n)$ gauge theory on fuzzy $S^2_N \times S^2_N$, 
which can be used to 
obtain any $\R^4_\theta$ as a scaling limit. The action is 
a simple generalization of the matrix model 
approach of \cite{Steinacker:2003sd} 
for fuzzy $S^2_N$. It differs from similar string-theoretical matrix models 
\cite{Ishibashi:1996xs} 
by adding a constraint-term, which ensures that the ``vacuum''
solution is stable and describes
the product of 2 spheres. The fluctuations of the covariant
coordinates then correspond as usual 
to the gauge fields, and the action reduces to 
ordinary Yang-Mills theory on $S^2 \times S^2$ in the limit $N \to
\infty$. The quantization of the model is defined by a finite integral
over the matrix degrees of freedom, which is shown to be convergent
due to the constraint term. We also give a  
gauge-fixed action with BRST symmetry. 

We then discuss some features of the model, in particular
a hidden $SO(6)$ invariance of the action which is broken explicitly 
by the constraint. This suggests some alternative formulations in
terms of ``collective matrices'', which are assembled from the
individual covariant coordinates (matrices). This turns
out to be very useful to construct a Dirac operator, and may help to
eventually study the quantization of the model. The stability of the model
without constraint is also discussed, and we show that the only 
flat directions of the $SO(6)$ -invariant action are
fluctuations of the constant radial modes of the 2 spheres. 

As a further test of the proposed gauge theory, we study 
in Section \ref{se: monopoles} topologically 
non-trivial solutions (instantons) on $S^2_N \times S^2_N$. 
We find in particular a simple class of solutions which can be interpreted as
$U(1)$ instantons with quantized flux, 
combined with a singular, localized ``flux tube''.
They are related to the 
so-called ``fluxon'' solutions
of $U(1)$ gauge theory on $\R^4_\theta$. Solutions which
can be interpreted as 2-dimensional spherical branes wrapping one of the two
spheres are also found and are matched with similar solutions on
$\R^4_\theta$.
We then study the relation of the model on $S^2_N \times S^2_N$ with
Yang-Mills theory on $\R^4_\theta$, and demonstrate
that the usual Yang-Mills action
on $\R^4_\theta$ is recovered in the appropriate scaling limit. 
Some aspects of $U(1)$ instantons
(``fluxons'') on  $\R^4_\theta$ are recalled 
in Section \ref{se: instantons}, and we show in detail
how they arise as limits of
the above non-trivial solutions on $S^2_N \times S^2_N$. In
particular, we are able to match the moduli space of $n$ fluxons,
corresponding to their location on $\R^4_\theta$ resp. 
$S^2_N \times S^2_N$. 
We find in particular that even though the field
strength in the ``bulk'' vanishes in the limit of $\R^4_\theta$, 
it does contribute to
the action on $S^2_N \times S^2_N$ with equal weight as the 
localized flux tube.  
This can be interpreted on $\R^4_\theta$ as a topological or
surface term at infinity.
Another unexpected feature on $S^2_N \times S^2_N$ is the appearance of 
certain ``superselection rules'', restricting the possible instanton numbers. 
In other words, not all instanton numbers on $\R^4_\theta$ are reproduced
for a given matrix size $\cN$, however they can be found by considering
matrices of different size. This
depends on the precise form of the constraint term in the
action, which is hence seen to imply also certain 
topological constraints.
To recover the full space of ADHM solutions on $\R^4_\theta$ 
starting from $S^2_N \times S^2_N$ remains an open challenge, which is
non-trivial since the concept of self-duality does not extend
naturally to the fuzzy case. 

We should mention here that topologically non-trivial
configurations have also been discussed more abstractly
in terms of projective modules
using a somewhat different 
formulation of gauge theory on fuzzy spaces,
see in particular \cite{Baez:1998he,Balachandran:1999hx}.% The precise
%relation between these formulations certainly deserves further study.  

Finally in Section \ref{sec:fermions} we include charged fermions 
in the fundamental representation of the gauge group, by giving a 
Dirac operator $\widehat{D}$
which in the large $N$ limit reduces to the
ordinary gauged Dirac operator on $S^2 \times S^2$. 
This Dirac operator is covariant under the $SO(6)$
symmetry of the embedding space $S^2 \times S^2 \subset \R^6$, and 
exactly anti-commutes with a chirality operator.
The 4-dimensional physical Dirac spinors are obtained by suitable
projections from 8-dimensional $SO(6)$ spinors.
This projection however commutes with $\widehat{D}$ only in the 
large $N$ limit, and is achieved by  giving one of the 2
spinors a large mass. Weyl spinors can then be defined using the
exact chirality operator.

\section{The fuzzy spaces  \protect\( S^{2}_{N}\protect \) 
and \protect\( S^{2}_{N_{L}}\times S^{2}_{N_{R}}\protect \) }

We start by recalling the definition of the fuzzy sphere in order to fix
our conventions and notation. The algebra of
functions on the fuzzy sphere is the finite algebra $S_N^2$
generated by Hermitian operators 
$x_i = (x_1, x_2, x_3)$ satisfying the defining relations
\bea 
[x_i, x_j] = i \Lambda_N \e_{ijk} x_k, \label{def1}\\
x_1^2 + x_2^2 +x_3^2  = R^2. \label{def2}
\eea
They are obtained from the \( N \)-dimensional representation of 
\( su(2) \) with generators \( \lambda _{i}\; (i=1,2,3) \)
and commutation relations 
\be
{}[\lambda _{i},\lambda _{j}]=i\epsilon _{ijk}\lambda _{k},
\qquad
\sum ^{3}_{i=1}\lambda _{i}\lambda _{i}=\frac{N^{2}-1}{4}
\label{lambda-algebra}
\ee
(see Appendix \ref{sec:useful}) by identifying 
\be
x_i = \Lambda_N\; \lambda_i, \qquad 
\Lambda_N = \frac{2R}{\sqrt{N^{2}-1}}. 
\ee
The noncommutativity parameter $\Lambda_N$ is of dimension length.
The algebra of functions $S^2_N$ therefore coincides with the simple
matrix algebra $Mat(N,\C)$. The normalized integral of a function 
$f \in S^2_N$ is given by the trace
\be
\int\limits_{S^2_N} f = \frac{4\pi  R^2}{N} \mathrm{tr}(f).
\ee 
The functions on the fuzzy sphere can be mapped to functions on
the commutative sphere $S^2$ using the decomposition into harmonics
under the action 
\be
J_i f = [\lambda_i,f]
\ee
of the rotation group $SU(2)$. One obtains analogs of the spherical
harmonics up to a maximal angular momentum $N-1$.
Therefore $S^2_N$ is a regularization of
$S^2$ with a UV cutoff, and the commutative sphere $S^2$ is recovered
in the limit $N \to \infty$.
Note also that for the standard representation
\eq{standard-rep}, 
entries in the upper-left block of the
matrices correspond to functions localized at $x_3 = R$.
In particular, the fuzzy delta-function at the ``north pole'' is
given by a suitably normalized projector of rank 1,
\be
\d^{(2)}_{NP}(x) = \frac{N}{4\pi R^2}\;|\frac{N-1}2\rangle\langle \frac{N-1}2|
\label{fuzzydelta}
\ee
where $|\frac{N-1}2\rangle$ is the highest weight state with
maximal eigenvalue of $\la_3$. 
Delta-functions with arbitrary
localization are obtained by rotating \eq{fuzzydelta}.

The simplest 4-dimensional generalization of the above is the
product $S^{2}_{N_{L}}\times S^{2}_{N_{R}}$ of 2 such fuzzy spheres,
with generally independent parameters $N_{L,R}$. It
is generated by a double set of representations
of \( su(2) \) commuting with each other, i. e. by 
\( \lambda ^{L}_{i},\lambda ^{R}_{i} \)
satisfying
\begin{eqnarray*}
{}[\lambda ^{L}_{i},\lambda ^{L}_{j}] &=&  i\epsilon _{ijk}\lambda
^{L}_{k},
\qquad
{}[\lambda ^{R}_{i},\lambda ^{R}_{j}] =  i\epsilon _{ijk}\lambda ^{R}_{k},\\
{}[\lambda ^{L}_{i},\lambda ^{R}_{j}] & = & 0
\end{eqnarray*}
for $i,j=1,2,3$, 
and  Casimirs
\beq
\sum ^{3}_{i=1}\lambda ^{L}_{i}\lambda ^{L}_{i}  =  \frac{N_{L}^{2}-1}{4},\qquad
\sum ^{3}_{i=1}\lambda ^{R}_{i}\lambda ^{R}_{i}  =  \frac{N_{R}^{2}-1}{4}.
\eeq
This can be realized as a tensor product of 2 fuzzy sphere algebras
\begin{eqnarray}
\lambda ^{L}_{i} & = & \lambda _{i}\otimes 1_{N_{R}\times N_{R},}\label{Lambda L} \\
\lambda _{i}^{R} & = & 1_{N_{L}\times N_{L}}\otimes \lambda _{i} \label{Lambda R},
\end{eqnarray}
hence as algebra we have 
$S^{2}_{N_{L}}\times S^{2}_{N_{R}} \cong \mathrm{Mat}(\cN,\C)$
where
\be
\cN = N_L N_R.
\ee
The normalized coordinate functions are given by
\be
x_i^{L,R} = \frac{2R}{\sqrt{(N^{L,R})^{2}-1}}\; \lambda_i^{L,R},
\qquad 
\sum (x_i^{L})^2 = R^2 = \sum (x_i^{R})^2 .
\ee
This space\footnote{In principle one could also introduce different 
radii $R^{L,R}$ for the 2
spheres, but for simplicity we will keep only one scale parameter $R$
(and usually we will set $R=1$).} 
can be viewed as regularization of 
$S^2 \times S^2 \subset \R^6$, and 
admits the symmetry group $SU(2)_L \times SU(2)_R \subset SO(6)$. 
The generators $x_i^{L,R}$
should be viewed as coordinates in an embedding space $\R^6$. 
The normalized integral of a function 
$f \in S^{2}_{N_{L}}\times S^{2}_{N_{R}}$ is now given by 
\be
\int\limits_{S^{2}_{N_{L}}\times S^{2}_{N_{R}}} f 
= \frac{16\pi^2  R^4}{\cN} \mathrm{tr}(f) =  \frac{V}{\cN} \mathrm{tr}(f), 
\label{int-4}
\ee 
where we define the volume $V:=16\pi^2  R^4$.
We will mainly consider $N_L = N_R$ in the following.

\subsection{\label{planelimit}The quantum plane limit $\R^4_\theta$}

It is well-known \cite{Chu:2001xi} that if a fuzzy sphere is blown up near a
given point, it can be used to obtain a (compactified) quantum plane:
Consider the tangential coordinates $x_{1,2}$ near the ``north
pole''. Setting
\be
R^2 = N\theta/2,
\label{R-theta}
\ee
they satisfy the commutation relations 
\be
[x_1, x_2]\; = \;i \frac{2R}N x_3 \;= \;i \frac{2R}N \sqrt{R^2-x_1^2-x_2^2} \;
=\; i\theta (1+ O(1/N)).
\ee
Therefore in  the large $N$ limit with \eq{R-theta} keeping $\theta$ fixed,
we recover\footnote{One could be more sophisticated and
  use the stereographic projections as in \cite{Chu:2001xi}, which
  leads essentially to the same results.}  
the commutation relation of the quantum plane,
\be
[x_1, x_2] = i \theta
\ee
up to corrections of order $\frac 1N$. Similarly,
starting with 
$S^{2}_{N_{L}}\times S^{2}_{N_{R}}$ and setting
\be
R^2 = N_{L,R} \theta_{L,R}/2,
\label{R-thetaLR}
\ee
we obtain in  the large $N_L, N_R$ limit
\bea
[x^L_i, x^L_j] &=& i\epsilon_{ij} \theta^L, \qquad
[x^R_i, x^R_j] = i\epsilon_{ij} \theta^R, \qquad\nn\\
\left[x^L_i, x^R_j\right] &=& 0 .
\eea
This is the most general form of $\R^4_\theta$ 
with coordinates $(x_1, ..., x_4) \equiv (x^L_1, x^L_2, x^R_1, x^R_2)$
(after a 
suitable orthogonal transformation).
The integral of a function $f(x)$ then becomes
\be
\int\limits_{S^{2}_{N_{L}}\times S^{2}_{N_{R}}} f(x) 
%= \frac{16\pi^2 R^4}{N_L N_R} Tr(f(x))
 \to 4\pi^2 \theta_L \theta_R  \mathrm{tr}(f(x)) =: \int\limits_{\R^4_\theta} f(x),
\ee 
which has indeed the standard normalization,
giving each ``Planck cell'' the appropriate volume.

\section{Gauge theory on fuzzy $S^{2}\times S^{2}$}

We start with the most general case, and 
construct a matrix model having \( S^{2}_{N_{L}}\times S^{2}_{N_{R}} \)
as its ground state. The fluctuations around this ground state will
produce a gauge theory. A simplified and more elegant formulation in
terms of ``collective matrices'' similar as in 
\cite{Steinacker:2003sd} for the fuzzy sphere will be given later in
Section \ref{so6 formulation}.

In the fuzzy case, it is natural to construct $S^2_L \times S^2_R$ as 
``submanifold'' of $\R^6$.
We therefore consider a multi-matrix model with
6 dynamical fields (``covariant coordinates'') \( B^{L}_{i} \) and
\( B_{i}^{R} \) \( (i=1,2,3) \), which are
$\cN \times \cN $ Hermitian matrices. As action we choose
the following generalization of the action in
\cite{Steinacker:2003sd,Presnajder:2003ak}, 
\be 
S  =  \frac{1}{g^{2}}
  \int \frac 12  F_{ia\: jb}  F_{ia\: jb}  +
\varphi_L^{2} + \varphi_R^{2} 
\label{action}
\ee
with \( a,b=L,R \) and \( i,j=1,2,3 \); 
summation over repeated indices is implied. 
Here $\varphi_{L,R}$ are
defined as
\be
\varphi_L := \frac 1{R^2} (B_{i}^{L}B_{i}^{L}-\frac{N_{L}^{2}-1}{4}), 
\qquad\varphi_R := \frac 1{R^2} (B_{i}^{R}B_{i}^{R}-\frac{N_{R}^{2}-1}{4}),
\label{phi-def}
\ee
and the terms $\;\varphi_L^{2} + \varphi_R^{2}\;$ in the action ensure
that the unwanted radial degrees of freedom are suppressed 
\cite{Steinacker:2003sd,Presnajder:2003ak}.
$R$ denotes the
radius of the two spheres, which we keep explicitly to have the
correct dimensions. 
The field strength is defined by
\begin{eqnarray}
F_{iL\: jL} &=& 
\frac 1{R^2}(i [B_{i}^{L},B_{j}^{L}]+ \epsilon _{ijk} B_{k}^{L}), \nn\\
F_{iR\: jR} &=& 
 \frac 1{R^2}(i [B_{i}^{R},B_{j}^{R}]+ \epsilon _{ijk} B_{k}^{R}), \nn\\
 F_{iL\: jR} &=&  \frac 1{R^2}(i [B_{i}^{L},B_{j}^{R}]).
\label{F-def}
\end{eqnarray}
This model \eq{action} is manifestly invariant under $SU(2)_L \times SU(2)_R$
rotations acting in the obvious way, 
and $U(\cN)$ gauge transformations
acting as $B^{L,R}_{i} \to U B^{L,R}_{i} U^{-1}$.
We will see below that this reduces indeed to the $U(1)$ Yang-Mills
action on $S^2 \times S^2$ in the commutative limit. 
Note that if the action \eq{action} is 
considered as a matrix model, the radius drops out using \eq{int-4}.
The equations of motion (e.o.m.) for \( B_{i}^{L} \) are
\begin{eqnarray}
&&{}\{B_{i}^{L},B_{j}^{L}B_{j}^{L}-\frac{N_{L}^{2}-1}{4}\}
  +(B_{i}^{L}+i\epsilon _{ijk}B_{j}^{L}B_{k}^{L}) \nn\\
&& +i\epsilon _{ijk}[B_{j}^{L},(B_{k}^{L}+i\epsilon_{krs}B_{r}^{L}B_{s}^{L})]
 +[B_{j}^{R},[B_{j}^{R},B_{i}^{L}]] =0  ,\label{EOM} 
\end{eqnarray}
and those for \( B_{i}^{R} \) are obtained by exchanging
\( L\leftrightarrow R \). 
By construction, the minimum or ground state of the action is given
by $F = \varphi =0$, hence
$B^{L,R}_{i} = \lambda _{i}^{L,R}$ as in (\ref{Lambda L}), (\ref{Lambda R})
up to gauge transformations;
cp. \cite{Grosse:2004wm} for a similar approach on $\C P^2$. We can therefore 
expand the ``covariant coordinates'' \( B_{i}^{L} \) and \( B_{i}^{R} \)
around the ground state
\be\label{B-A-relation}
B^{a}_{i}  = \lambda _{i}^{a}+R{A}_{i}^{a},
\ee
where  $a\in\{L,R\}$ and ${A}_{i}^{a}$ is small. 
Then ${A}_{i}^{L,R}$ transforms under gauge transformations as
\be\label{gaugetrafo-A}
{A}_{i}^{L,R} \to {A'}_{i}^{L,R} 
=   U {A}_{i}^{L,R} U^{-1} + U [{\la}_{i}^{L,R},U^{-1}],
\ee
and the field strength takes a more familiar form\footnote{We do not
  distinguish between upper and lower indices $L,R$.},
\begin{eqnarray}
F_{iL\: jL} &=& 
i([\frac{\lambda^{L}_{i}}R, A^{L}_{j}]
 -[\frac{\lambda^{L}_{j}}{R}, A^{L}_{i}]
 +[A^{L}_{i},A^{L}_{j}]),\nn\\
F_{iR\: jR} &=& 
i([\frac{\lambda^{R}_{i}}R, A^{R}_{j}]
 -[\frac{\lambda^{R}_{j}}{R},A^{R}_{i}]
 +[ A^{R}_{i}, A^{R}_{j}]),\nn\\
 F_{iL\: jR} &=& i([\frac{\lambda^{L}_{i}}R, A^{R}_{j}]
 -[\frac{\lambda^{R}_{j}}{R}, A^{L}_{i}]
 +[ A^{L}_{i},A^{R}_{j}]).
\label{F-A}
\end{eqnarray}
So far, the spheres are  described in terms of 3 Cartesian 
covariant coordinates each. In the commutative limit, we can
separate the radial and tangential degrees of freedom.
There are many ways to do this; 
perhaps the most elegant for the
present purpose is to note that
the terms $\int \varphi_L^2 + \varphi_R^2$
in the action  imply that
$\varphi_{L,R}$ is bounded
for configurations with finite action. Using
\be
\varphi_L  =\; \frac {\lambda _{i}^{L}}R {A}_{i}^{L} 
+ {A}_{i}^{L} \frac{\lambda _{i}^{L}}R 
+ {A}_{i}^{L}{A}_{i}^{L},
\label{phi-2}
\ee
and similarly for $\varphi_R$ it follows that
\be
x_i {A}_{i}^{a} + {A}_{i}^{a} x_i = O(\frac{\varphi}N)
\label{A-constraint}
\ee
for finite $A_i^a$. This means that ${A}_{i}^{a}$ is tangential in the
(commutative) large $N$ limit. 
Alternatively, one could consider 
$\phi_L = N \varphi_L$,
which would acquire a mass of order $N$ and decouple from the
other fields\footnote{The  constraints 
$\varphi_L = 0 = \varphi_R$ could also be imposed by hand;
however the suppression through the above terms in the action is more
flexible, as we will see in Section \ref{se: monopoles}.}.
The commutative limit of (\ref{action}) therefore gives 
the standard action for electrodynamics on \( S^{2}\times S^{2} \),
\[
S=\frac{1}{2g^{2}}\int\limits_{S^2\times S^2} F_{ia\: jb}^t  F_{ia\: jb}^t
\]
with \( a,b=L,R \) and \( i,j=1,2,3 \). Here
\( F_{iL\: jR}^t\) denotes the usual tangential field strength.
This can be seen most easily noting that e.g. at the north pole
$x_3^{L,R} = R$, one can replace
\be
i[\frac{\lambda_{i}^{L,R}}R,\cdot ] \;\;\rightarrow \;\;
-\varepsilon_{ij}\frac{\partial }{\partial x^{L,R}_{j}} 
\ee
in the commutative limit,
so that upon identifying the commutative gauge fields $A^{(cl)}_i$
via 
\be
A^{(cl) L,R}_i = -\varepsilon_{ij} A^{L,R}_i
\ee
the field strength is  given by the standard expression
\( F_{iL\: jR}^t =\pp i^{L}A_{j}^{(cl)R}-\pp j^{R}A_{i}^{(cl)L} \) etc.

\subsection*{$U(k)$ gauge theory}

The above action generalizes immediately to the nonabelian case,
keeping precisely the same action \eq{action}, \eq{phi-def} but
replacing the matrices $B_i^{L,R}$ by $k \cN \times k \cN $ 
matrices, cp. 
\cite{Steinacker:2003sd}. 
Expanding them in terms of (generalized) Gell-Mann matrices, the same
action \eq{action}  is the fuzzy version of nonabelian
$U(k)$ Yang-Mills on $S^2 \times S^2$.

\section{A formulation based on  $SO(6)$}
\label{so6 formulation}

The above action can be cast into a nicer
form by assembling the matrices $B_i^{L,R}$ into bigger ``collective
matrices'', following \cite{Steinacker:2003sd}.
Since it is natural from the fuzzy point of view
to embed $S^2 \times S^2 \subset R^6$ with corresponding embedding of
the symmetry group $SO(3)_L \times SO(3)_R \subset SO(6)$, we consider 
\be
B_\mu = (B_i^L, B_i^R) 
\ee
(Greek indices \( \mu,\nu   \) denoting from now on all the six
dimensions)
to be the $6$ -dimensional irrep  of $so(6) \cong su(4)$. 
Since $(4) \tens (4) = (6) \oplus (10)$, 
it is natural to
introduce the intertwiners 
\be
\g_\mu = (\g_i^L,\g_i^R) = (\g_\mu)^{\a, \b}
\ee
of $(6) \subset (4) \tens (4)$.
where $\a,\b$ denote indices of
$(4)$. 
We could then assemble our
dynamical fields into a single $4\cN \times 4 \cN $ matrix
\be
B =  B_\mu \g_\mu \;\; + \mathrm{const}\cdot  \one.
\label{B-def}
\ee
Of course the most general such $4\cN \times 4 \cN $ matrix 
contains far too many degrees of freedom, and we have to constrain 
these $B$ further. Since $SU(4)$ acts on $B$ as $B \to U^T B U$,
the $\g_\mu$ can be chosen as totally anti-symmetric matrices, which
precisely singles out the $(6) \subset (4) \tens (4)$.
One can moreover impose
\be
(\g_i^L)^\dagger = \g_i^L, \qquad (\g_i^R)^\dagger = -\g_i^R,
\label{g-conj}
\ee
and
\begin{eqnarray}
\gamma^{L}_{i}\gamma^{L}_{j} & = & \delta_{ij}+i\epsilon_{ijk}\gamma^{L}_{k},
\label{g-rel-L}\\
\gamma^{R}_{i}\gamma_{j}^{R} & = &-\delta_{ij}-\epsilon_{ijk}\gamma_{k}^{R},
\label{g-rel-R}\\
{}[\gamma_{i}^{L},\gamma_{j}^{R}] & = & 0,
\label{g-L-R}
\end{eqnarray}
which will be assumed from now on;
we will give two explicit such representations in
\eq{gamma-rep-1}, \eq{gamma-wolfg}.
This would suggest to constrain $B$ to be antisymmetric. 
However, the component fields $B_\mu$ are naturally considered as
Hermitian rather than symmetric matrices. 
Furthermore, since the $\g_\mu = (\g_\mu)^{\a, \b}$ 
have two upper indices, they do not form
an algebra. There are now 2 ways to proceed. We can either separate them
again by introducing two $4\cN \times 4 \cN $ matrices,
\be
B^L = \frac 12 + B_i^L \g_i^L, \qquad B^R = \frac i2 + B_i^R \g_i^R,
\label{B-L-R}
\ee
breaking $SO(6) \to SO(3) \times SO(3)$. This will be pursued in
Appendix \ref{sec:4Nmatrices}.
Alternatively, we can use the $\g_\mu$ with the above properties 
to construct the $8 \times 8$ Gamma-matrices 
\begin{equation}
\label{Gamma}
\Gamma ^{\mu }=\left( \begin{array}{cc}
0 & \gamma ^{\mu }\\
\gamma ^{\mu \dagger } & 0
\end{array}\right) ,
\end{equation}
which generate the \( SO(6) \)-Clifford algebra
\be
\{\Gamma ^{\mu },\Gamma ^{\nu }\}=\left( \begin{array}{cc}
\gamma ^{\mu }\gamma ^{\nu \dagger }+\gamma ^{\nu }\gamma ^{\mu \dagger } & 0\\
0 & \gamma ^{\mu \dagger }\gamma ^{\nu }+\gamma ^{\nu \dagger }\gamma ^{\mu }
\end{array}\right) =2\delta ^{\mu \nu }.
\label{clifford-so6}
\ee
This suggests to consider the single Hermitian \( 8\cN \times 8\cN \) matrix
\be
C=\Gamma^{\mu }B_{\mu }+ C_0 
 = \left( \begin{array}{cc}
0 & B^L \\
B^L & 0
\end{array}\right) + \left( \begin{array}{cc}
0 &  B^R\\
- B^R & 0
\end{array}\right) =: C^L + C^R ,
\label{C-def}
\ee
where
$C_0 = C_0^L + C_0^{R}$ denote the constant 
\( 8\times 8 \)-matrices 
\begin{eqnarray}
C_0^L & = & -\frac i2 \G^L_1 \G^L_2\G^L_3 
= \frac{1}{2}\left( \begin{array}{cc}
0 & 1\\
1 & 0
\end{array}\right) ,\label{B0-l}
\\
C_0^{R} & = &  -\frac i2 \G^R_1 \G^R_2\G^R_3 
= \frac{i}{2}\left( \begin{array}{cc}
0 & 1\\
-1 & 0
\end{array}\right) \label{B0-R}
\end{eqnarray}
in the above basis.
This is very close to the approach in \cite{Steinacker:2003sd}, and
using the Clifford algebra and the above definitions one finds indeed
\be
C^{2} =B_{\mu }B_{\mu }+\frac{1}{2}+\Sigma _{8}^{\mu \nu }F_{\mu \nu }.
\label{c-2}
\ee
Here $\Sigma _{8}^{\mu\nu } = -\frac i4 [\G_\mu,\G_\nu]$, and 
 the field strength $F_{\mu \nu }$ coincides with the definition
in \eq{F-def} if written in the $L-R$ notation,
\[
F_{ia\: jb}=i[B_{ia},B_{jb}] + \delta _{ab}\epsilon _{ijk}B_{ka}.
\] 
Therefore  the action 
\be
S_{6}= \mathrm{Tr}((C^{2}-\frac{N^{2}}{2})^{2})
= 8  \mathrm{tr} (B_{\mu }B_{\mu}-\frac{N^{2}-1}{2})^{2}
+ 4  \mathrm{tr}  F_{\mu \nu }F_{\mu \nu }
\label{action-so6}
\ee
is quite close to what we want. The only difference is the
term \( (B_{\mu }B_{\mu }-\frac{N^{2}-1}{2})^{2} \) instead of 
\((B_{iL}B_{iL}-\frac{N_L^{2}-1}{4})^{2}
+(B_{iR}B_{iR}-\frac{N_R^{2}-1}{4})^{2}\), for $2 N^2 = N_L^2 + N_R^2$.
This difference is easy to understand: since \eq{action-so6}
is \( SO(6) \)-invariant,
the ground state should be some \( S^{5} \). 
We therefore have to break this $SO(6)$- invariance explicitly, 
which will be done in the next section. However before doing that, 
let us try to understand action (\ref{action-so6}) better
and see whether it leads to a meaningful 4-dimensional field theory. 
We show in Appendix \ref{app:stability} by carefully integrating out the
scalar components of $B_i^{L,R}$ 
that the  $SO(6)$- invariant constraint term in (\ref{action-so6}) 
induces the second term in the following effective action
\be
S_{6}^{\mathrm{eff}} \sim 4  \mathrm{tr} \left( F_{\mu \nu }F_{\mu \nu }
 - (F_{iL}x_{iL}-F_{iR}x_{iR})
\frac{1}{4(\frac{1}{2}-\dd _{\mu }\dd_{\mu
  })}(F_{iL}x_{iL}-F_{iR}x_{iR})
\right)
\ee
in the commutative limit, where $F_{iL} = \frac 12 \epsilon_{ijk} F_{jL\, kL}$
etc.
Comparing the second term with \( F_{\mu\nu  }F_{\mu\nu } \), we see
that the zero mode of the Laplace operator $\dd _{\mu }\dd_{\mu }$
can produce a contribution
that cancels the corresponding contribution from $F_{\mu\nu  }F_{\mu\nu }$, 
but that all higher modes are smaller
by at least a factor of $2(\frac{1}{2}-\dd _{\mu }\dd _{\mu })$.
Therefore, the action (\ref{action-so6}) is positive definite
except for the obvious zero mode 
$\d B_i^L =\epsilon,\;\; \d B_i^R =-\epsilon$. 
This means that the geometry of \( S^{2}_L\times S^{2}_R \) is locally
stable even with the \( SO(6) \)-symmetry unbroken, except for
opposite fluctuations of the radii.

\subsection{Breaking $SO(6) \to SO(3) \times SO(3)$}

To obtain the original action \eq{action} for \( S^{2}\times S^{2} \),
we have to break the \( SO(6) \)-symmetry down to \( SO(3)\times SO(3) \).
We can do this by using the left and right gauge fields $C^L$ and
$C^R$ introduced in \eq{C-def} 
separately.
Their squares are
\begin{eqnarray*}
C_{L}^{2} & = & B_{iL}B_{iL}+\frac{1}{4}+\left( \begin{array}{cc}
\gamma ^{i}_{L} & 0\\
0 & \gamma ^{i}_{L}
\end{array}\right) (B_{iL}+i\epsilon _{ijk}B_{jL}B_{kL}),\\
C^{2}_{R} & = & B_{iR}B_{iR}+\frac{1}{4}-i\: \left( \begin{array}{cc}
\gamma ^{i}_{R} & 0\\
0 & \gamma ^{i}_{R}
\end{array}\right) (B_{iR}+i\epsilon _{ijk}B_{jR}B_{kR}).
\end{eqnarray*}
As both \( \gamma ^{i}_{L},\: \gamma _{R}^{i} \) and \( \gamma ^{i}_{L}\gamma ^{j}_{R} \)
are traceless, we have
\[
S_{\mathrm{break}}:= \mathrm{Tr} ((C^{2}_{L}-\frac{N_L^{2}}{4})
(C_{R}^{2}-\frac{N_R^{2}}{4}))
=8 \mathrm{Tr} ((B_{iL}B_{iL}-\frac{N_L^{2}-1}{4})(B_{iR}B_{iR}-\frac{N_R^{2}-1}{4})).
\]
With these terms we can recover our action as
\begin{eqnarray}
S &=& S_{6}-2 S_{\mathrm{break}}  
   = \mathrm{Tr} \big((C^{2}-\frac{N^{2}}{2})^{2}-\{C^{2}_{L}-\frac{N_L^{2}}{4},C_{R}^{2}-\frac{N_R^{2}}{4}\}\big)\nonumber \\
 &=& 8\,  \mathrm{tr}\,
 \big((B_{iL}B_{iL}-\frac{N_L^{2}-1}{4})^{2}+(B_{iR}B_{iR}-\frac{N_R^{2}-1}{4})^{2}+\frac{1}{2}F_{\mu \nu }F_{\mu \nu }\big),
\label{eq: action S_6 - S_break} 
\end{eqnarray}
which is precisely the action \eq{action} for 
gauge theory on \( S^{2}_{N_L}\times S^{2}_{N_R} \) 
omitting the overall constants. Hence the action is formulated as
2-matrix model, however with highly constrained matrices $C_L,C_R$.
This formulation using the Gamma-matrices is very natural
and useful if one wants
to couple the gauge fields to fermions, as discussed in Section 
\ref{sec:fermions}.

For simplicity, we will only consider $N_L = N_R = N$ from now on.

\section{Quantization}

The quantization of the gauge theory defined by \eq{action} 
or its reformulation
\eq{eq: action S_6 - S_break} is straightforward in principle, by 
a ``path integral'' over the Hermitian matrices
\beq
Z[J] = \int d  B_\mu e^{-S[B_\mu] + \mathrm{tr} B_\mu J_\mu}.
\label{ZJ}
\eeq
Note that there is no need to fix the gauge since the
gauge group $U(\cN)$ is compact.
The above path integral is well-defined and finite for 
any fixed $\cN$. To see this, it is enough to show that the 
integral
\(\int d  B_\mu \exp(-(B^{L}_{i}B^{L}_{i}-(N^{2}-1)/4)^{2} -
(B^{R}_{i}B^{R}_{i}-(N^{2}-1)/4)^{2})\) converges, since the
 contributions from the field strength further suppress the integrand.
This integral is obviously convergent for any fixed $N$.

For perturbative
computations it is necessary to fix the gauge, and to substitute
gauge invariance by BRST-invariance. Such a gauge-fixed action will be
presented next.

\subsection{BRST Symmetry}

To construct a gauge-fixed BRST-invariant action, we have
to introduce ghost fields \( c \) and anti-ghost fields \( \bar{c} \).
These are fermionic fields, more precisely
\(\cN\times \cN- \) matrices with entries which are Grassman
variables. 

The full gauge-fixed action reads:
\[
S_{\mathrm{BRST}}=S +\frac 1{\cN} \mathrm{tr}\big(\bar{c}[\lambda _{\mu },[B_{\mu },c]]
  -(\frac{\alpha }{2} b-[\lambda _{\mu },B_{\mu}])b\big)\, ,
\]
 where \( b \) is an auxiliary (Nakanishi-Lautrup)
field. This action is invariant with respect to the following 
BRST-transformations:
\begin{eqnarray}
sB_{\mu }=[B_{\mu },c] &  & sc=cc\\
s\bar{c}= b &  & s b=0
\label{eq: BRST-trafos} 
\end{eqnarray}
 (matrix product is understood), where the BRST-differential \( s \)
acts on a product of fields as follows:
\[
s(XY)=X(sY)+(-1)^{\varepsilon _{Y}}(sX)Y\, .
\]
 Here \( \varepsilon _{Y} \) denotes the Grassman-parity of \( Y \)
\[
\varepsilon _{Y}=\left\{ \begin{array}{cc}
0 & Y\, \textrm{bosonic}\\
1 & Y\, \textrm{fermionic}\, .
\end{array}\right. 
\]
 As usual, it is not difficult to check that these BRST-transformations
are indeed nilpotent, i.e. 
\[
s^{2}=0\, .
\]
Integrating out the auxiliary field \( b \) leads to the
following action 
\[
S'_{\mathrm{BRST}}=S +\frac 1{\cN} \mathrm{tr}\big(\bar{c}[\lambda _{\mu },[B_{\mu
},c]]-\frac{1}{2\alpha }[\lambda _{\mu },B_{\mu }][\lambda _{\nu
},B_{\nu }]\big)\, .
\]
 Setting \( \alpha =1 \) corresponds to the Feynman gauge. 
This is indeed what one would
obtain by the Faddeev-Popov procedure.
The action
\( S' \) is invariant with respect to the following operations:
\begin{eqnarray*}
s'B_{\mu } & = & [B_{\mu },c]\\
s'c & = & cc\\
s'\bar{c} & = & [\lambda _{\mu },B_{\mu }]\, .
\end{eqnarray*}
 Since we have used the equations of motion of \( b \),
the BRST-differential \( s' \) is \emph{not} nilpotent off-shell
anymore but still we have 
\[
s'^{2}|_{\textrm{on}-\textrm{shell}}=0\, .
\]

\section{Topologically non-trivial solutions on 
\( S^{2}_{N}\times S^{2}_{N}\protect \)}
\label{se: monopoles}

In order to understand better the non-trivial solutions found below, 
we first note that the classical space $S^2 \times S^2$ is symplectic
with symplectic form 
\be
\omega = \omega^L + \omega^R,
\ee
where 
\be
\omega^{L} = \frac 1{4 \pi R^3} \epsilon_{i j k} 
  x^{L}_i dx^{L}_j dx^{L}_k
\ee
and similarly $\omega^{R}$.
The normalization is chosen such that
\be
\int_{S^2_{L,R}} \omega^{L,R} = 1 = \int_{S^2\times S^2}\omega^L
\wedge \omega^R
\ee
so that $\omega^{L},\omega^{R}$ generate the integer cohomology 
$H^{*}(S^2\times S^2,\Z)$. Noting that $\omega$ is self-dual while 
$\tilde \omega := \omega^L - \omega^R$ is anti-selfdual, it follows
immediately that both $F = 2\pi \omega$ and $F = 2\pi\tilde \omega$ are
solutions of the Abelian
field equations. More generally, any
\be
F^{(m_L,m_R)} = 2\pi m_L \omega^L + 2 \pi m_R \omega^R
\ee
for any integers $m_L, m_R$ is a solution. 
In bundle language, they correspond to 
products of 2 monopole bundles with connections and 
monopole number $m_{L,R}$ over $S^2_{L,R}$.
Following the literature we will denote any such non-trivial solution 
as instanton.

\subsection{Instantons and fluxons}

We are interested in similar non-trivial solutions of the
 e.o.m. (\ref{EOM}) in the fuzzy case.  The monopole solutions
on the  fuzzy
 sphere $S^2_N$ are given by \reps $\lambda_i^{N-m}$
of $su(2)$ of size $N-m$ \cite{Karabali:2001te}, which 
lead to the classical monopole gauge fields in the commutative limit 
as shown in \cite{Steinacker:2003sd}. It is hence 
easy to guess that we will obtain solutions on $S^2_N\times S^2_N$
by taking  products of these:
\begin{eqnarray}\label{monopoles L}
B_{i}^{L} & = & \alpha ^{L}\; \lambda_{i}^{N-m_{L}}\otimes \one_{N-m_{R}},\\
\label{monopoles R}B_{i}^{R} & = & \alpha ^{R}\; \one_{N-m_{L}}\otimes \lambda_{i}^{N-m_{R}}
\label{monopole-ansatz}
\end{eqnarray}
where  \( \lambda _{i}^{N-m_{L,R}} \) are the \( N-m_{L,R} \) dimensional
generators of $su(2)$. It is 
not difficult to verify that these are solutions of
(\ref{EOM}) with \( \alpha ^{L,R}=1+\frac{m_{L,R}}{N} \) for \( m_{L,R}\ll N
\), with field strength  
\be
F_{iLjL} = - \frac{m^{L}}{2 R^3} \epsilon_{ijk} x_k^{L}, 
\quad F_{iRjR} = - \frac{m^{R}}{2 R^3} \epsilon_{ijk} x_k^{R}, 
\quad F_{iLjR} =0,
\label{monopole-field}
\ee
while $B\cdot B - \frac{N^2-1}4 \to 0$ as $N \to \infty$.
This means that 
$F = - 2\pi m^{L}\omega^{L} - 2\pi m^{R}\omega^{R}$ 
in the commutative limit, so that indeed
\be
\int\limits_{S_2^{L,R}} \frac{F}{2\pi} = -  m^{L,R}.
\label{monopole-flux}
\ee
Notice that the Ansatz \eq{monopole-ansatz} 
implies that all matrices have size 
$\cN = (N-m_{L})(N-m_{R})$, which is inconsistent if we require that 
$\cN = N^2$ in order to
have the original $S^{2}_{N}\times S^{2}_{N}$ vacuum.
Therefore it appears that these solutions live in a different configuration
space, similar as the commutative monopoles which live on different
bundles. 
However, the situation is in fact more interesting: the above
solutions can be embedded in the {\em same} configuration
spaces of $N^2 \times N^2$ matrices as the vacuum solution if we
combine them with other solutions, which have
finite action in four dimensions\footnote{as opposed to 2 dimensions, which is
the reason why they were not considered in \cite{Steinacker:2003sd}}. 
They are in fact crucial to recover some of the known 
$U(1)$ instantons in the limit 
\(S^2_N \to \mathbb {R}^{2}_{\theta} \) resp. 
\(S^2_N\times S^2_N \to \mathbb {R}^{4}_{\theta} \), as we will see. 
Consider the following Ansatz 
\[
B_{i}^{L,R}=\mathrm{diag}(d_{i,1}^{L,R},...,d^{L,R}_{i,n})
\]
in terms of  diagonal matrices (ignoring the size of the matrices for
the moment).
These are solutions of (\ref{EOM}) in two cases,
\be
\sum_i d^{L,R}_{i,k}d^{L,R}_{i,k}=
\left\{\begin{array}{ll} \frac{N^{2}-3}{4}, & \mbox{type A}\\
                        0 , & \mbox{type B}
                \end{array}\right.
\label{singular-solutionsAB}
\ee
(i.e. $ d^{L,R}_{i,k} =0$ in type B).
The associated field strength is
\be
F_{iLjL} = \frac{\epsilon_{ijk}}{R^2}\; 
\mathrm{diag}(d_{k,1}^{L},...,d^{L}_{k,n}), 
\quad F_{LR} =0,
\ee
and a similar formula for $F_{iRjR}$. The constraint term is then
$(B\cdot B - \frac{N^2-1}4) \to -\frac 12$ for type A, and
$(B\cdot B - \frac{N^2-1}4) \to-\frac{N^2-1}4$ for type B 
in the large $N$ limit. 
In particular, only the type A solutions 
will have a finite contribution 
\be
S_{\mathrm{fluxon}} =  \frac{V}{g^2 \cN} \left(\frac n{4R^4} 
   + \frac{2n}{R^4}\frac{N^{2}-3}{4}\right) 
  \;\;\to  \frac{8\pi^2}{g^2}\; n
\label{defect-action}
\ee
to the action\footnote{A finite
  action can also be obtained for the type $B$ solution
  using a slightly modified action \eq{action-2}, as discussed below.},
which for $N \to \infty$ is only due to the field strength. 
We will see below that 
these type A solutions can be interpreted as a localized flux or vortex,
and we will call them ``fluxons'' since they will reduce in a certain
scaling limit to 
solutions on $\R^4_\theta$ which are sometimes denoted as such 
\cite{Polychronakos:2000zm,Gross:2000ph,Harvey:2000jb}.

One can now combine these ``fluxon'' solutions with the monopole
solutions \eq{monopole-ansatz} in the form
\begin{eqnarray}\label{combined solution}
B_{i}^{L} & = & \left( \begin{array}{cc}
\alpha ^{L}\; \lambda _{i}^{N-m_{L}}\tens\one_{N-m_{R}} & 0\\
0 & \mathrm{diag}(d_{i,1}^{L},...,d^{L}_{i,n})
\end{array}\right) ,\nn \\
B_{i}^{R} & = & \left( \begin{array}{cc}
\alpha ^{R}\; \one_{N-m_{L}}\otimes \lambda _{i}^{N-m_{R}} & 0\\
0 & \mathrm{diag}(d_{i,1}^{R},...,d^{R}_{i,n}) 
\end{array}\right).
\label{eq: R instantons on S2 S2} 
\end{eqnarray}
These are now matrices of size $\cN = (N-m_{L})(N-m_{R})+n$, which must
agree with $\cN = N^2$, say. This is clearly possible for 
\be
m_L = -m_R = m, \quad n = m^2,
\label{m-constraints}
\ee
while for $m_L \neq -m_R$ the contribution from the fluxons 
would be infinite since $n = O(N)$.
To understand these solutions, we can compute the gauge field from
(\ref{B-A-relation}), 
\be
\cA^L_i = \frac 1R\left( B_i^L  
- \la_i^N \tens\one_N \right) = \cA^L_i(x^L,x^R).
\ee
To evaluate this, we first have to choose a gauge, i.e. a unitary
transformation $U$
for \eq{eq: R instantons on S2 S2} which allows to express e.g.
$\la_i^{N-m_L} \tens\one_{N-m_R}$ in terms of  
$x_i^L \propto\la_i^N \tens\one_N$ and $x_i^R \propto \one_N \tens\la_i^N$.
For example, in the case $m_L = -m_R = m$ this can be done using
a unitary map
\be
U:\;\; \C^{N-m} \tens \C^{N+m} \oplus \C^{m^2} \to \C^{N} \tens \C^{N},
\ee
mapping a $(N-m) \times (N+m)$ matrix into a $N \times N$ matrix by 
trivially 
matching the upper-left corner in the obvious way, and fitting $\C^{m^2}$
into the remaining lower-right corner. 
With this being understood, one can write
\bea
R\cA^L_i(x^L,x^R) &=& (\a^L \la_i^{N-m} - \la_i^{N}) \tens\one_{N+m}  
 + \la_i^{N} \tens(\one_{N+m} - \one_{N}) +(d-\mathrm{terms}) \nn\\
 &=& A_i^{(m)}(x^L) \;\;+ \;\mathrm{sing}(x_3^L = -R, x_3^R =-R)
\eea 
where $A_i^{(m)}(x^L)$ is indeed 
the gauge field of a monopole with charge
$m$ on $S^2_L$ in the large $N$ limit, as was checked explicitly in 
\cite{Steinacker:2003sd}. Here $\mathrm{sing}(x_3^L = -R, x_3^R =-R)$ 
indicates a field localized at
the ``south pole'' of $S^2_L$ and/or $S^2_R$ which becomes singular 
for large $N$. It originates both from
``cutting and pasting'' the bottom and right border 
of the above matrices using $U$ (leading to
singular gauge fields but regular field strength at the south poles), 
as well as the $d$-block (leading to a singular field strength).
To see this recall that in general for the standard representation
\eq{standard-rep} of fuzzy spheres, 
entries in the lower-right block of the
matrices correspond to functions localized at $x_3 = -R$, cp. \eq{fuzzydelta}.
The gauge field near this singularity will be
studied in more detail in Section \ref{sec:fluxon-limit}.
The  field strength is
\be
F_{iLjL}  = - \frac{m^{L}}{2 R^3} \epsilon_{ijk} x_k^{L}
+ \epsilon_{ijk} \frac{1}{R^2}\; \sum_{i=1}^n\; d_{k,i}^{L} P_i 
\label{eq: F instantons on S2 S2} 
\ee
in the commutative limit, where 
$P_i $ are projectors in the algebra of functions on 
$S^2_N \times S^2_N$ of rank $1$; recalling \eq{fuzzydelta}, 
they should be interpreted as 
delta-functions $P_i =  \frac{V}{N^2}\; \d^{(4)}(x_3=-R)$.
Similar formulae hold for $\cA^L_i(x^L,x^R)$ and $F_{iRjR}$, while
$F_{LR} =0$.

We assumed above that these delta-functions are
localized at the south poles $x_3^L=x_3^R=-R$.
However, the location of these delta-functions 
can be chosen freely using gauge transformations. 
This can be seen by applying suitable 
successive gauge transformations using $N-k$-dimensional 
irreps of $SU(2)$ for $k=0,1,...,m-1$, 
which from the classical point of view all
correspond to global rotations, successively moving the individual delta-peaks.
Therefore the solution \eq{eq: R instantons on S2 S2} should 
in general be
interpreted as monopole on $S^2 \times S^2$ with monopole number
$m_L = -m_R = m$, 
combined with a localized singular field strength characterized by 
its position and a vector $ d_{k,i}^{L}$. 
We will see in Section \ref{sec:scaling} 
that it becomes the ``fluxon'' solution in the planar limit  $\R^4_\theta$; 
we therefore also call it a ``fluxon''.

The total action of these solutions \eq{eq: R instantons on S2 S2}
is the sum of the 
contributions from the monopole field plus the contribution from the
fluxons \eq{defect-action}, which both give the same contribution
\be\label{final action}
S_{(m)} = \frac{4\pi^2}{g^2} \left(2m^2 + 2 m^2\right)
\label{action-fluxon-total}
\ee
in the large $N$ limit, using \eq{m-constraints}. 
The first term is due to the global monopole field 
\eq{monopole-field}, and 
the second term is the contribution of the fluxons through the
localized field strength.

The interpretation of these solutions depends on the scaling limit $N \to
\infty$ which we want to consider. We have seen that 
in the commutative limit keeping
$R = \mathrm{const}$, these solutions become  commutative monopoles on 
$S^2 \times S^2$ with magnetic charges
$m_L = - m_R$, plus additional localized ``fluxon''
degrees of freedom. For large $R$, the field strength of the 
monopoles vanishes, leaving only the localized fluxons. In particular, 
we will see in the following section
that in the scaling limit $S^2_N \times S^2_N \to \R^4_\theta$
only the fluxons survive and become well-known
solutions for gauge theory on $\R^4_\theta$. Away from this localized 
fluxon the gauge field becomes a flat connection, which  is however
topologically nontrivial. 
This is very interesting as it shows that one can indeed use these
fuzzy spaces as regularization for gauge theory on
\(\mathbb {R}^{2n}_{\theta} \).

A final remark is in order:
if we fix the size $\cN$ of the matrices, only
certain fluxon  and monopole numbers are allowed, given by \eq{m-constraints}.
Otherwise the number $n$ of fluxons  and hence the action 
would diverge with $N$. This can be seen as an interesting feature of
our model: viewed as a regularization of gauge theory 
on $\R^4_\theta$, this points to possible subtleties of
defining the admissible field
configurations in infinite-dimensional Hilbert spaces and relations
with topological terms in the action. 
On the other hand, we could accommodate the most general
solutions including also type B solutions \eq{singular-solutionsAB} by
modifying the action  similar as in \cite{Steinacker:2003sd}. For example,
\bea
S  =  \frac{1}{g^{2}}
  \int\Big( \frac{4 B^{L}_{i}B^{L}_{i}}{N^2 R^4}
    (B^{L}_{i}B^{L}_{i}-\frac{N^{2}_{L}-1}{4})^{2}
    +\frac{4B^{R}_{i}B^{R}_{i}}{R^4}
    (B^{R}_{i}B^{R}_{i}-\frac{N^{2}_{R}-1}{4})^{2} % \nn\\
 + \frac 12  F_{ia, jb}  F_{ia,jb} \Big) \nn\\
\label{action-2}
\eea
leads to the same commutative action, but with a  
vanishing action for the Dirac string in the type B solutions.

\subsection{Spherical branes}

Consider the following solutions
\begin{eqnarray}
B_{i}^{L} & = & \left( \begin{array}{cc}
\alpha ^{L}\; \lambda _{i}^{N-m} & 0\\
0 & \mathrm{diag}(d_{i,1},...,d_{i,m})
\end{array}\right)\tens \one_{N} ,\nn \\
B_{i}^{R} & = & \one_{N} \tens \lambda _{i}^{N}
\label{S2-branes} 
\end{eqnarray}
which are matrices of size $\cN = N^2$. The corresponding field
strength is 
\bea
\quad F_{iLjL} & = &  - \frac{m}{2 R^3} \epsilon_{ijk} x_k^{L}
   +  \epsilon_{ijk} \frac{1}{R^2}\; \sum_{i=1}^m\; d_{k,i} P_i  \nn\\
F_{RR} &=& F_{LR} =0
\eea
where $P_i$ are projectors in the algebra of functions on 
$S^2_L$ of rank $1$ which should be interpreted as 
delta-functions $P_i = \frac{4\pi R^2}{N}\; \d^{(2)}(x_3=-R)$.
In particular the gauge field  $A$ vanishes on $S^2_R$, while  
on $S^2_L$ there is a monopole field together with 
a singularity at a point. This is similar to the
 fluxons on the previous section, but now only on $S^2_L$. 
This leads to the interpretation as 
2-dimensional brane  wrapping on
$S^2_R$, located at a point on $S^2_L$.
The action for these solutions is infinite.
In the limit $S^2_N \times S^2_N \to \R^4_\theta$, the flux will be
located at a 2-dimensional hyperplane. 
Such solutions for gauge theory on $\R^4_\theta$ were found in 
\cite{Aganagic:2000mh,Gross:2000ss}, which would be recovered in the scaling limit 
$S^2_N \times S^2_N \to \R^4_\theta$ as discussed in Section \ref{sec:scaling}.
In a similar way, we can interpret solutions with any $m_L, m_R$ as
branes wrapping on $S^2_L$ and $S^2_R$.

\section{Gauge theory on 
\protect\( \mathbb {R}^{4}_{\theta }\protect \) from
\protect\( S^{2}_{N_{L}}\times S^{2}_{N_{R}}\protect \)}
\label{sec:scaling}

We saw in Section \ref{planelimit} that \( \mathbb {R}^{4}_{\theta }
\)
can be obtained as a scaling limit of fuzzy 
\( S^{2}_{N_{L}}\times S^{2}_{N_{R}} \). Here we will extend
this scaling also to the covariant coordinates \( B_{\mu } \),
thereby relating the gauge theory on \( S^{2}_{N_{L}}\times
S^{2}_{N_{R}} \) to 
that on \( \mathbb {R}^{4}_{\theta } \) and hence providing 
a regularization for the latter. We will in
particular relate the instanton solutions on these two spaces.

On noncommutative
 \( \mathbb {R}^{2}_{\theta } \), 
all $U(1)$-instantons 
were constructed and classified in \cite{Gross:2000ss}. 
They can be interpreted as localized flux solutions, sometimes called
fluxons. 
One can indeed recover these
instantons from corresponding solutions on $S^{2}_{N}$, 
as we will show below. However
since we are mainly interested in the 4-dimensional
case here, we will only present the corresponding constructions 
on \( S^{2}_{N_{L}}\times S^{2}_{N_{R}} \) resp. $\mathbb {R}^{4}_{\theta }$
here, without discussing the 2-dimensional case separately. It can be
recovered in an obvious way from the considerations below.

The situation on $\R^4_\theta$ is more complicated, and
there are different types of non-trivial 
$U(1)$ ``instanton'' solutions on $\mathbb {R}^{4}_{\theta }$. 
Assuming that $\theta_{\mu \nu}$ is self-dual, there are two types of 
instantons: first, there exist  straightforward
generalizations of the localized ``fluxon'' solutions with self-dual
field strength. These will be discussed in detail here, and we will
show how these solutions can be recovered as scaling limits of 
the solutions  \eq{eq: R instantons on S2 S2}
on \(S^{2}_{N_{L}}\times S^{2}_{N_{R}} \). 
This is one of the main results of the present
paper. In particular, the moduli of the fluxon solutions on 
$\mathbb {R}^{4}_{\theta }$ will be related to the free parameters
$d_i^{L,R}$ in \eq{eq: R instantons on S2 S2}. This supports our suggestion to use gauge theory on 
\(S^{2}_{N_{L}}\times S^{2}_{N_{R}} \) as a regularization for 
gauge theory on $\mathbb {R}^{4}_{\theta }$. However there are other 
types of $U(1)$ instantons on $\mathbb {R}^{4}_{\theta }$ which were
found through a noncommutative version of the ADHM equations
\cite{Nekrasov:1998ss,Furuuchi:1999kv,Chu:2001cx,Hamanaka:2001dr,Ivanova:2005fh,Wimmer:2005bz},
in
particular anti-selfdual instantons which are much less localized than
the fluxon solutions. To find the corresponding 
solutions on \(S^{2}_{N_{L}}\times S^{2}_{N_{R}} \)
is an interesting open challenge.

\subsection{The action }
\label{sec:action-scaling}

The most general noncommutative $\mathbb {R}^{4}_{\theta }$ 
is generated by the coordinates
subject to the commutation relations 
\be
[x_{\mu},x_{\nu}]=i\theta_{\mu\nu}\,,
\label{x-CR}
\ee
 where  $\mu,\nu\in\{1,\dots,4\}$. 
Using suitable rotations, 
$\theta_{\mu\nu}$ can always be cast in the following form:
\[
\theta_{\mu \nu}=\left(\begin{array}{cccc}
0 & \theta_{12} & 0 & 0\\
-\theta_{12} & 0 & 0 & 0\\
0 & 0 & 0 & \theta_{34}\\
0 & 0 & -\theta_{34} & 0\end{array}\right)\:.\]
We will assume that  \( \theta_{12}>0 \)
and \( \theta_{34}>0 \) for simplicity in this section.
Then define
\begin{eqnarray}
X_{1,2} & := & \sqrt{\frac{2\theta _{12}}{N_{L}}}\; B_{1,2}^{L}\;,
\label{eq: scaling limit1} \\
X_{3,4} & := &  \sqrt{\frac{2\theta _{34}}{N_{R}}}\; B_{1,2}^{R}\;,
\label{eq: scaling limit2}\\
\phi ^{L,R} & := & B^{L,R}_{3}-\frac{N_{L,R}}{2}+\frac{1}{N_{L,R}}((B_{1}^{L,R})^{2}+(B_{2}^{L,R})^{2})\;,\label{eq: scaling limit3}
\end{eqnarray}
which should be interpreted as a blow-up near the north pole.
In the scaling limit \eq{R-thetaLR},
\be
R^{2}=\frac{1}{2}N_{L}\theta _{34}=\frac{1}{2}N_{R}\theta _{12} 
\quad\to\;\;\infty
\label{R-thetaLR-2}
\ee
the \( X \) will become the covariant coordinates on 
the ``tangential'' \( \mathbb {R}^{4}_{\theta } \)
as \( N_{L,R}\rightarrow \infty  \), and \( \phi  \) remains
an auxiliary field. To see this, we compute for the field strength 
\begin{eqnarray*}
\frac{1}{R^{2}}([B_{1}^{L},B^{R}_{1}]) & = & \frac{1}{\theta
  _{12}\theta _{34}}[X_{1},X_{3}], \quad\mbox{etc.,}\\
\frac{1}{R^{2}}(B^{L}_{1}+i[B^{L}_{2},B^{L}_{3}]) & = & \sqrt{\frac{1}{\theta _{12}\theta _{34}R^{2}}}\;(X_{1}+i[X_{2},\phi ^{L}]-\frac{i}{2\theta _{12}}[X_{2},(X_{1})^{2}])\\
\frac{1}{R^{2}}(B^{L}_{2}+i[B^{L}_{3},B^{L}_{1}]) & = & \sqrt{\frac{1}{\theta _{12}\theta _{34}R^{2}}}\;(X_{2}+i[X_{1},\phi ^{L}]-\frac{i}{2\theta _{12}}[X_{1},(X_{2})^{2}])\\
\frac{1}{R^{2}}(B^{L}_{3}+i[B^{L}_{1},B^{L}_{2}]) & = & \frac{1}{\theta _{12}\theta _{34}}(\theta _{12}+i[X_{1},X_{2}]+\frac{\theta _{12}\theta _{34}}{R^{2}}\phi _{L}-\frac{\theta _{12}\theta _{34}^{2}}{2R^{4}}((X_{1})^{2}+(X_{2})^{2}))\, .
\end{eqnarray*}
Analogous expressions hold for \( B^{R}_{i} \). For the potential
term we get
\bea
\frac{1}{R^{2}}(B^{L}_{i}B^{L}_{i}-\frac{N_{L}^{2}-1}{4}) &=&
\frac{1}{\theta _{34}}\phi ^{L}
+\frac{2}{R^{2}}((\phi^{L})^{2}+\frac{1}{4})
 -\frac{1}{\theta _{12}R^{2}}\{\phi^{L},(X_{1})^{2}+(X_{2})^{2}\} \nn\\
&& +\frac{1}{\theta^{2}_{12}R^{2}}((X_{1})^{2}+(X_{2})^{2})^{2}. \nn
\eea
 We immediately see that the only terms from action (\ref{action})
involving \( \phi ^{L,R} \) are
\[
\frac{1}{\theta _{34}^{2}}(\phi ^{L})^{2}+\frac{1}{\theta
  _{12}^{2}}(\phi ^{R})^{2}+ O(\frac{1}{R}),
\]
and therefore we can integrate them out in the limit 
\( R\rightarrow \infty  \).
In the leading order in \( R \)
 the remaining terms give the standard
action 
\[
S=-\frac{1}{2g^{2}\theta _{12}^{2}\theta _{34}^{2}}\int
([X_{\mu},X_{\nu}]-i\theta _{\mu\nu})^{2}
\]
for a gauge theory on \( \mathbb {R}^{4}_{\theta } \) for general
$\theta_{\mu\nu}$.
The $X_{\mu}$ are interpreted as 
``covariant coordinates'', which can be written as\footnote{We do not 
distinguish between upper and lower indices.}
\[
X_{\mu}:=x_{\mu}+i \theta_{\mu \nu} A_{\nu}.
\]
Hence the gauge fields $A_{\mu}$
describe the fluctuations around the vacuum. In particular, note that 
our regularization procedure clearly fixes the rank of the gauge
group, unlike in the naive definition on $\R^d_\theta$ as discussed in
\cite{Gross:2000ss}. The generalization to the $U(n)$ case is obvious.

\subsection{$U(1)$ Instantons on \protect\( \mathbb {R}^{4}_{\theta }\protect
  \)}
\label{se: instantons}

The construction of instanton solutions for the two-dimensional noncommutative
plane given in \cite{Gross:2000ss} can be easily generalized to the
four-dimensional case. We shall recall and discuss these 
4-dimensional ``fluxon'' solutions 
in some detail here, in order to understand the relation with the
above solutions.
To simplify the following formulas, 
we restrict our discussion from now on to the selfdual case
\[
\theta_{\mu\nu}=
\frac{1}{2}\varepsilon_{\mu\nu\rho\sigma}\theta_{\rho\sigma}
\]
and denote 
\[
\theta:=\theta_{12} = \theta_{34}; 
\]
the generalizations to the antiselfdual and the general case are
obvious. Then
the action for $U(1)$ gauge theory 
on $\mathbb{R}_{\theta}^{4}$ reads
\be \label{eq: action on R4}
S = \frac{(2\pi)^2}{2 g^2\theta^2}\;\mathrm{tr} (F_{\mu\nu} F_{\mu\nu})
\ee
where
\be
F_{\mu\nu} =i([X_{\mu},X_{\nu}]-i\theta_{\mu\nu})
\ee
is the field strength.
In terms of the complex coordinates
\begin{eqnarray*}
x_{\pm L}:=x_{1}\pm ix_{2} & , & x_{\pm R}:=x_{3}\pm ix_{4},
\end{eqnarray*}
the commutation relations \eq{x-CR} take the form 
\be \label{algebra}
[x_{+a},x_{-b}]=2\theta\delta_{ab},\,\,\,[x_{+a},x_{+b}]=[x_{-a},x_{-b}]=0,
\ee 
where $a,b\in \{L,R\}$.
The Fock-space representation $\mathcal{H}$
of (\ref{algebra}) has the standard basis
\[
|n_{1},n_{2}\rangle, \quad n_1,n_2 \in \mathbb{N}\,,
\]
with 
\begin{eqnarray*}
x_{-L}|n_{1},n_{2}\rangle=\sqrt{2\theta}\sqrt{n_{1}+1}|n_{1}+1,n_{2}\rangle, & x_{+L}|n_{1},n_{2}\rangle=\sqrt{2\theta}\sqrt{n_{1}}|n_{1}-1,n_{2}\rangle\\
x_{-R}|n_{1},n_{2}\rangle=\sqrt{2\theta}\sqrt{n_{2}+1}|n_{1},n_{2}+1\rangle,
&
x_{+R}|n_{1},n_{2}\rangle=\sqrt{2\theta}\sqrt{n_{2}}|n_{1},n_{2}-1\rangle\,.
\end{eqnarray*}
Similarly, using the complex covariant coordinates $X_{\pm a}$ 
\begin{eqnarray}
X_{\pm L}=X_1 \pm i X_2  & ,  & X_{\pm R}=X_3 \pm i X_4
\end{eqnarray}
and the corresponding field strength
\[
F_{\alpha a, \beta b}=[X_{\alpha a},X_{\beta b}]
 -2 \theta \varepsilon_{\alpha\beta}\delta_{ab}\,
\]
with $a,b\in\{L,R\}$ and $\alpha,\beta \in\{+,-\}$,
the action (\ref{eq: action on R4}) can be written in the
form 
\[
S= \frac{\pi^{2}}{g^2\theta^{2}}\;\mathrm{tr}(\sum_{a}F_{+a,-a}F_{+a,-a}-\sum_{a,b}F_{+a,+b}F_{-a,-b}).
\]
Then the equations of motion are given by:
\begin{equation}
\sum_{a,\alpha}[X_{\alpha a},(F_{\alpha a,\beta b})^{\dagger}]=0\,.\label{eq:
  EOM of plane}
\end{equation}
Let us consider a finite dimensional subvectorspace $V_{n}$ of $\mathcal{H}$
of dimension $n$, which we can assume (using a unitary gauge transformation)
to be spanned by a finite set of vectors
$|n_{1},n_{2}\rangle\in\mathcal{H}$, 
\begin{equation}
V_{n}=\langle\{|i_k,j_k\rangle;\;\; k = 1,...,n\}\rangle\,.\label{eq: subspace of
  H}
\end{equation}
Following  \cite{Gross:2000ss} one finds  solutions
to the equations of motion given by\footnote{Note that $[X^{(n)}_{+L},X^{(n)}_{+R}]=[X^{(n)}_{+L},X^{(n)}_{-R}]=[X^{(n)}_{-L},X^{(n)}_{+R}]=[X^{(n)}_{-L},X^{(n)}_{-R}]=0$
.} 
\begin{eqnarray}
X^{(n)}_{+L} & := & Sx_{+L}S^{\dagger}+\sum_{k=1}^n\gamma_{k}^{L}
|i_k,j_k\rangle\langle i_k,j_k|
\label{eq: sol for  instantons on R4 L}\\
X^{(n)}_{+R} & := &
Sx_{+R}S^{\dagger}+\sum_{k=1}^n \gamma_{k}^{R}
|i_k,j_k\rangle\langle i_k,j_k|\,.
\label{eq: sol for instantons on R4 R}
\end{eqnarray}
Here $\gamma_{k}^{L,R} \in \C$ determine the position of the fluxons, and
$S$ denotes a partial isometry from $\mathcal{H}$ to
 $\mathcal{H}$\textbackslash{}$V_{n}$ with 
$S^\dagger S = \one,\; S S^\dagger = \one - P_{_{V_{n}}}$, where 
\[
P_{_{V_{n}}}:=\sum_{k=1}^n| i_k,j_k\rangle\langle  i_k,j_k|
\]
 is the projection operator onto the subspace $V_{n}$.
The field strength $F_{\mu\nu}$ for this solution is
\[
F_{\mu\nu}= P_{_{V_{n}}}\theta_{\mu\nu}.
\]
In particular, the
action corresponding to the instanton solution 
(\ref{eq: sol for instantons on R4 L},\ref{eq: sol for instantons on R4 R})
is proportional to the dimension of the subspace $V_{n}$ 
\[
S[X^{(n)}_{\pm a}]=\frac{8 \pi^2}{g^2}\; \mathrm{tr}(P_{_{V_{n}}}) 
 =\frac{8 \pi^2}{g^2}\; n .
\]
We will see in the next section that this class of solutions can be
reproduced by instanton solutions \eq{eq: R instantons on S2 S2}
on $S_{N_{L}}^{2}\times S_{N_{R}}^{2}$ in a suitable scaling limit.
Let us stress again that this is only one class of $U(1)$-instanton solutions
for $\mathbb{R}_{\theta}^{4}$ which is called ``fluxons'',
since they can be interpreted as localized flux. 
The localization can be seen as follows: recall \cite{Furuuchi:2000vc} that
the above projection operators can be represented on the space of commutative
functions (using a normal-ordering prescription) as
\[
|k^{1},k^{2}\rangle\langle k^{1},k^{2}|\cong
\frac{1}{k^1!k^2!}
(\frac{x^{-L}}{\sqrt{2\theta}})^{k^{1}}(\frac{x^{+L}}{\sqrt{2\theta}})^{k^{1}}
(\frac{x^{-R}}{\sqrt{2\theta}})^{k^{2}}(\frac{x^{+R}}{\sqrt{2\theta}})^{k^{2}}
e^{-\frac{x^{+L}x^{-L}}{2\theta}-\frac{x^{+R}x^{-R}}{2\theta}}\,.
\]
Hence the above field strengths $F_{\mu\nu}= P_{_{V_{n}}}\theta_{\mu\nu}$ are
superpositions of Gauss-functions which are localized in a 
region in space of size $\sqrt{\theta}$.

\subsection{Instantons on $\mathbb
  {R}^{4}_{\theta }$ from $S^{2}_{N}\times S^{2}_{N}$}
\label{sec:fluxon-limit}

With the scaling limit of Section \ref{sec:action-scaling},
 the gauge theory on  \( S^{2}_{N}\times S^{2}_{N} \)
provides us with a regularization for the gauge theory on \( \mathbb {R}^{4}_{\theta } \).
Of course, such a regularization might affect the topological features
of the theory, an effect we want to investigate in this section. For
this, we will map the topologically nontrivial solutions found in
Section \ref{se: monopoles} on \( S^{2}_{N}\times S^{2}_{N} \) to \( \mathbb {R}^{4}_{\theta } \). 

Consider again the solutions (\ref{combined solution}) that combine
the fluxon solutions with the monopoles, with the fluxons at the
north pole instead of the south pole because we want to study their
structure.  
Their scaling
limit as in \eq{eq: scaling limit1} gives
\begin{eqnarray}
X_{i} & = & \sqrt{\frac{2\theta }{N}}\left( \begin{array}{cc}
\mathrm{diag}(d_{i,1}^{L},...,d^{L}_{i,n})  & 0\\
0 & \alpha ^{L}\;  \lambda _{i}^{N-m}\otimes \one
\end{array}\right) ,\label{scaled fluxons L} \\
X_{i+2} & = & \sqrt{\frac{2\theta }{N}}\left( \begin{array}{cc}
\mathrm{diag}(d_{i,1}^{R},...,d^{R}_{i,n}) & 0\\
0 & \alpha ^{R}\; \one\otimes \lambda _{i}^{N+m}
\end{array}\right) \label{scaled fluxons R} 
\end{eqnarray}
for $i=1,2$.
Recalling that the rescaled 
\( \lambda_{1,2}  \) on \( S^{2}_{N_{L}}\times S^{2}_{N_{R}} \)
become the \( x_{\pm} \)'s on \( \mathbb {R}^{4}_{\theta } \) 
in the scaling limit
\[
\sqrt{\frac{2\theta}{N}}(\lambda^{L,R}_{1}\pm i\lambda
_{2}^{L,R})\rightarrow x_{\pm L,R},
\]
we see that 
(\ref{scaled fluxons L}) and (\ref{scaled fluxons R})
become the instantons (\ref{eq: sol for  instantons on R4 L},
\ref{eq: sol for  instantons on R4 R}) on \( \mathbb {R}^{4}_{\theta } \),
\begin{eqnarray}
X_{1}+ iX_{2} & \rightarrow  & X_{+ L}^{(n)}=Sx_{+ L}S^{\dagger}
+\sum_{k=1}^n\gamma_{k}^{L}|i_k,j_k\rangle\langle i_k,j_k|,
\label{limit to instantons 1} \\
X_{3}+iX_{4} & \rightarrow  & X^{(n)}_{+ R}=Sx_{+ R}S^{\dagger }+
\sum_{k=1}^n\gamma_{k}^{R}|i_k,j_k\rangle\langle i_k,j_k|.
\label{limit to instantons 2} 
\end{eqnarray}
Here the $(d_i)$-block acting on a basis $|i_{k},j_{k}\rangle$ 
of $V_{n} \subset \mathcal{H} \cong \C^{\cN}$ becomes the
projector part of (\ref{limit to instantons 1},
\ref{limit to instantons 2}) with 
\begin{eqnarray}
\sqrt{\frac{2\theta}{N}}\;d^{L,R}_{1,k} & \rightarrow  & 
\mathrm{Re}\gamma ^{L,R}_{k},\nn\\
\sqrt{\frac{2\theta}{N}}\;d^{L,R}_{2,k} & \rightarrow  & 
\mathrm{Im}\gamma ^{L,R}_{k},
\label{d-gamma-relation}
\end{eqnarray}
and the monopole block becomes  \( Sx_+ S^{\dagger } \)
where $S$ is a partial isometry  from $\mathcal{H}$ to
 $\mathcal{H}$\textbackslash{}$V_{n}$.  
Note that we can recover any value for the \( \gamma  \)'s in this
scaling, solving the constraint \( d_{i}d_{i}=\frac{N^{2}-3}{4} \) 
for \( d_{3}\sim \frac{N}{2} \). 
Therefore the full moduli space of the fluxon solutions 
(\ref{eq: sol for  instantons on R4 L},
\ref{eq: sol for  instantons on R4 R}) on \( \mathbb {R}^{4}_{\theta }
\) can be recovered in this way.
Furthermore, the meaning of the parameters $\g^{L,R}$ is easy to
understand in our approach: Note first that using a rotation 
(which acts also on the indices) followed
by a gauge transformation, the $d_i$ can be fixed to be radial at the
north pole, $d_{i}^{L,R} \sim(0,0,N/2)$. This is a fluxon localized at
the north pole. Now apply a ``translation'' at the north pole, 
which corresponds
to a suitable rotation on the sphere. Rotating
the vector $d_i^{L,R}$ in the scaling limit 
amounts to a translation of the 
$\gamma^{L,R}_{k}$ according to \eq{d-gamma-relation}, 
which therefore parametrize the position of the fluxons.

It has been noted \cite{Douglas:2001ba} that the \( Sx_+ S^{\dagger }
\)
correspond
to a pure (but topologically nontrivial) gauge, which can qualitatively
be seen already in two dimensions. There, the isomorphism \( S:|k\rangle\rightarrow |k+n\rangle \) is
basically \( (\frac{x_{-}}{\sqrt{x_{-}x_+}})^{n}\sim (\frac{x-iy}{r})^{n}\sim e^{in\varphi } \)
and therefore the gauge field \( A_{i}=S\dd _{i}S^{\dagger } \)
has a winding number \( n \). The topological nature of the \( Sx_{+}S^{\dagger } \)
is even more evident in our setting, as they are the limit of the
monopole solutions (\ref{monopoles L}, \ref{monopoles R}) on 
\( S^{2}_{N}\times S^{2}_{N} \). Moreover, note that
their contribution to the action (\ref{final action}) survives
the scaling: even though the field strength
vanishes as $R \to \infty$, 
the integral gives a finite contribution equal to the contribution
of the fluxon part. This topological {}``surface term'' is usually
omitted in the literature on \( \mathbb {R}^{4}_{\theta } \), but
becomes apparent in the regularized theory.

So it seems that we recovered all the instantons of Section \ref{se: instantons}, but
in fact there is an important detail that we haven't discussed jet.
It is the embedding of the \( n \)-dimensional fluxons and the \( (N-m)(N+m) \)-dimensional
monopole solutions into the \( N^{2} \)-dimensional matrices of the ground
state. Such an embedding is clearly only possible for \( n=m^{2} \).
This means that the regularized theory has some kind of  
``superselection rule'' for
the dimension of the allowed instantons, a rule that did not exist
in the unregularized theory\footnote{Note that this is different 
in two dimensions. There, a rank \(n\) fluxon can be combined 
with a \((N-n)\)-dimensional monopole block and all the instantons on 
\( \mathbb R^2_{\theta}\) can be recovered. Furthermore, the actions 
for the fluxons and the monopoles scale differently with \(N\). 
Therefore the action for the monopoles vanishes in the scaling
 limit that produces a gauge theory on \( \mathbb R^2_{\theta}\) 
with rescaled coupling constant.}. 

One way to allow arbitrary instanton numbers is to allow the size
$\cN$ of the matrices to vary. However, this is less satisfactory 
as it destroys the unification of topological sectors which is 
a beautiful feature of noncommutative gauge theory. 
On the other hand, the type B solutions (\ref{singular-solutionsAB})
together with the changed action (\ref{action-2}) might 
allow the construction of the missing instantons. The idea is
to fill up the unnecessary \( m^{2}-n \) places with \( d_{i}=0 \).
The changed action would not suppress such solutions any more, and
in fact they would not even contribute to the action. 
This amounts to adding a discrete sector to the theory which
accommodates these type B solutions, but decouples from the rest of
the model. Whether or not one wants to do this appears to be a matter
of choice. This emphasizes again the importance of a careful
regularization of the theory. 
It would be very interesting to see what happens in other
regularizations e.g. using gauge theory on noncommutative tori 
 or fuzzy $\C P^2$.

\section{Fermions}
\label{sec:fermions}

\subsection{The commutative Dirac operator on \protect\( S^{2}\times S^{2}\protect \)}

To find a form of the commutative Dirac operator on 
\( S^{2}\times S^{2} \) which is suitable for the fuzzy case,
one can generalize the approach of \cite{Grosse:1995pr} for
$S^2$, which is carried out in detail in Appendix \ref{sec:so6-dirac-4}:
One can write the flat \( SO(6) \) Dirac
operator \( D_{6} \) in 2 different forms, 
using spherical coordinates of the
spheres and also using the usual flat Euclidean coordinates. Then
one can relate $D_6$ with the curved four-dimensional Dirac operator
\( D_{4} \) on \( S^{2}\times S^{2} \) in the same spherical coordinates.
This leads to an explicit expression for \( D_{4} \)
involving only the angular momentum generators, which is easy to generalize
to the fuzzy case. 
The result is rather obvious and
easy to guess:
\begin{equation}
\label{Dirac operator}
D_4=\Gamma ^{\mu }J_{\mu }+\left( \begin{array}{cc}
0 & 1\\
1 & 0
\end{array}\right) +i\left( \begin{array}{cc}
0 & 1\\
-1 & 0
\end{array}\right)  = \Gamma ^{\mu }J_{\mu } + 2 C_0,
\end{equation}
which is  clearly a $SO(3) \times SO(3)$-covariant Hermitian 
first-oder differential operator. Here 
$\Gamma ^{\mu} $ generate the $SO(6)$ Clifford algebra 
\eq{clifford-so6}, $C_0$ is defined in \eq{B0-R},
and we put $R=1$ for simplicity here.
However this Dirac operator is reducible, acting on 
8-dimensional spinors  \( \Psi _{8} \) corresponding 
to the \(SO(6)\) Clifford algebra.
Hence $\Psi _{8}$ should be a combination
of two independent 4-component Dirac spinors on
the 4-dimensional space $S^2 \times S^2$. To see this, we will construct
explicit projectors projecting onto these 4-dimensional spinors, 
and identify the appropriate 4-dimensional chirality operators.
This will provide us with the desired physical Dirac or Weyl
fermions.

\subsubsection{Chirality and projections for the spinors}
\label{chapter on projections} 

There are 3 obvious operators which anti-commute with $D_4$. One
is the usual 6-dimensional chirality operator
\be 
\G := i \G^L_1 \G^L_2 \G^L_3 \G^R_1 \G^R_2 \G^R_3
  = \left( \begin{array}{cc}
-1 & 0\\
0 & 1
\end{array}\right),
\label{chiral-G-op}
\ee
which satisfies 
\be
\{D_4,\G\}= 0, \qquad \G^\dagger = \G, \qquad\G^2 = 1.
\label{G-chiral}
\ee
The 8-component spinors $\Psi _{8}$ split
accordingly  into two 4-component spinors
$\Psi_8 = \left(\begin{array}{c}\psi_\a \\ \obar\psi_{\obar{\b}}\end{array}\right)$,
which transform as $(4)$ resp. $(\obar{4})$ under $so(6) \cong su(4)$;
recall the related discussion in Section \ref{so6 formulation}.
The other operators of interest are 
\[
\chi _{L}=\Gamma^{iL}x_{iL}\; \; \; \mbox {and}
\; \; \; \chi _{R}=\Gamma ^{iR}x_{iR}.
\]
They preserve $SO(3)\times SO(3) \subset SO(6)$, and satisfy
\[
\{D_4,\chi _{L,R}\}= 0 = \{\chi _{L},\chi _{R}\}
\]
as well as
\[
\chi_{L,R} ^{2} =1.
\]
We will also use
\begin{equation}
\label{chiralityop}
\chi =\frac{1}{\sqrt{2}}\; \Gamma ^{\mu }x_{\mu } 
= \frac{1}{\sqrt{2}}\;(\chi_L + \chi_R)
\end{equation}
which satisfies similar relations. 
This means that
\begin{equation}
\label{projector}
P_{\pm }=\frac{1}{2}(1\pm i\chi _{L}\chi _{R})
\end{equation}
 with
\begin{equation}
\label{properties of p+-}
P_{\pm }^{2}=P_{\pm },\; \; \; \; P_{+}+P_{-}=1\; \; \; \; \mbox {and}\; \; \; \; P_{+}P_{-}=0
\end{equation}
are Hermitian projectors commuting with the Dirac 
operator on \( S^{2}\times S^{2} \) as well as with $\G$,
\be
P^{\dagger }_{\pm }=P_{\pm }\; \; \; \; \mbox {and}\; \; \; \; 
[P_{\pm },D_{4}]= [P_{\pm },\G] = 0.
\label{P-properties}
\ee
 Therefore they project onto subspaces which are preserved by
$D_4$ and $\G$, and are invariant under $SO(3) \times SO(3)$. 
Hence the spinor Lagrangian can be written as
\[
\Psi ^{\dagger }_{8}D_4\Psi _{8}=\Psi ^{\dagger }_{+}D_4\Psi _{+}
+\Psi^{\dagger }_{-}D_4\Psi _{-}
\]
involving two Dirac spinors $\Psi _{\pm} = P_\pm \Psi_8$.
In order to get one 4-component Dirac spinor, we can e.g. impose the constraint
\begin{equation}
\label{constraint}
P_{+}\Psi _{8}=\Psi _{8},
\end{equation}
or equivalently give one of the two components a large mass,
by adding a term
\be
M_- \Psi^{\dagger }_{8}P_-\Psi_8
\ee
to the action with $M_- \to \infty$. 
The physical chirality operator is now identified using 
\eq{P-properties} and \eq{G-chiral}
as $\G$ acting on $\Psi_+$.
It can be used to define 2-component Weyl spinors on $S^2 \times S^2$.

To make the above more explicit, consider
the north-pole of the spheres, i.e. 
\[
x_{L}=\left( \begin{array}{c}
1\\
0\\
0
\end{array}\right) \; \; \; \; \mbox {and}\; \; \; \; x_{R}=\left( \begin{array}{c}
1\\
0\\
0
\end{array}\right).
\]
In the basis (\ref{Gamma}) for the Clifford algebra we then get explicitly
\[
P_{\pm }=\frac{1}{2}(1\pm i\left( \begin{array}{cc}
-\gamma ^{1}_{L}\gamma ^{1}_{R} & 0\\
0 & \gamma ^{1}_{L}\gamma ^{1}_{R}
\end{array}\right) )=\frac{1}{2}(1\pm \sigma _{3}\otimes \sigma
_{3}\otimes \sigma _{3}).
\]
 This means that
\[
P_{+}=\mathrm{diag}(1,0,0,1,0,1,1,0)
\]
projects onto a 4-dimensional subspace exactly as expected.

\subsection{Gauged fuzzy Dirac and chirality operators}

To find fuzzy analogues of (\ref{Dirac operator}) and
(\ref{chiralityop})
coupled to the gauge fields,
we recall the connection between the gauge theory on \( S^{2}\times S^{2} \)
and the \( SO(6) \) Gamma matrices established in Section 
\ref{so6 formulation}. In
the spirit of that section a natural fuzzy spinor action would involve
\begin{equation}
\label{try}
\Psi ^{\dagger }C\Psi ,
\end{equation}
 where \( \Psi  \) is now a \( 8 \cN\times \cN \)-matrix 
(with Grassman entries). 
Of course, (\ref{try}) does not have the appropriate commutative limit, 
but we can split \( C \) into
a fuzzy Dirac operator \( \widehat{D} \) and the operator
$\widehat{\chi }$ defined by
\begin{equation}
\label{fuzzy chirality operator}
\widehat{\chi }\Psi =\frac{\sqrt{2}}{N}\;
(\Gamma ^{\mu }\Psi \lambda _{\mu }-C_0\Psi ),
\end{equation}
which generalizes \eq{chiralityop}; we used here the definition 
\eq{B0-l},\eq{B0-R} of $C_0$.
This operator satisfies
\[
\widehat{\chi }^{2}=1,
\]
and reduces to (\ref{chiralityop}) in the commutative limit. Note also
that $\widehat{\chi }$ commutes with gauge transformations,
since the coordinates \( \lambda _{\mu } \) are acting from the right
in \eq{fuzzy chirality operator}.
Setting 
\[
\widehat{J}_{\mu }\Psi =[\lambda _{\mu },\Psi ],
\]
we get for the fuzzy Dirac operator 
\begin{equation}
\label{fuzzy dirac operator}
\widehat{D}=C-\frac{N}{\sqrt{2}}
 \;\widehat{\chi }
=\Gamma^{\mu }(\widehat{J}_{\mu }+  A_{\mu })+2C_0
 = \Gamma^{\mu }\widehat{\mathcal{D}}_{\mu} +2C_0.
\end{equation}
Here\footnote{We set $R=1$ in this section for simplicity.}
\begin{equation}
\widehat{\mathcal{D}}_{\mu}:=\widehat{J}_{\mu} +  A_{\mu} 
\end{equation}
is a covariant derivative operator,
i.e. $U\widehat{\mathcal{D}}_{\mu}\psi = \widehat{\mathcal{D}}'_{\mu}U
\psi$ which is easily verified using (\ref{gaugetrafo-A}).
This $\widehat{D}$ clearly has the correct classical limit 
(\ref{Dirac operator}) for vanishing $A$, 
and the gauge fields are coupled correctly.
In particular, this definition of $\widehat{D}$
 applies also to the topologically non-trivial solutions
of Section \ref{se: monopoles} without any modifications.
Moreover, the chirality operator $\G$ as defined in \eq{chiral-G-op} 
anti-commutes with $\widehat{D}$ also in the fuzzy case,
\be
\{\widehat{D}, \G\} =0.
\ee
In particular there is no need to consider e.g. fuzzy Ginsparg-Wilson
operators as in the 2-dimensional case 
\cite{Balachandran:2003ay,Ydri:2002nt,Aoki:2002fq}.
However, the anticommutator of $\widehat{D}$ and $\widehat{\chi}$ 
no longer vanishes. We find
\be
\{\widehat{D},\widehat{\chi }\}  =  -\frac{\sqrt{2}}{N}\;\Big(2(\lambda _{\mu }+ A_{\mu })\widehat{J}_{\mu }-2 A_{\mu }\lambda _{\mu }
+ \{\Gamma^{\mu},C_0\} \widehat{\mathcal{D}}_{\mu}
+2 \Big)
= O(\frac 1N), 
\ee
since \( x_{\mu }J_{\mu }= O(\frac 1N) \) and 
\( x_{\mu }A_{\mu }= O(\frac 1N)\) 
using \eq{A-constraint}.
Furthermore, using some identities given at the beginning of Section 
\ref{so6 formulation} we obtain for $\widehat{D}^2 \psi$: 
\begin{eqnarray}
\widehat{D}^2 \psi  & = & (\Sigma^{\mu\nu}F_{\mu\nu}+
\widehat{\mathcal{D}}_{\mu}\widehat{\mathcal{D}}_{\mu}  
+ \{\Gamma^{\mu},C_0\} \widehat{\mathcal{D}}_{\mu} 
%+ \{\widehat{D},C_0\}
+ 2)\psi \nonumber \\
 & =: & (\Sigma^{\mu\nu}F_{\mu\nu} + \widehat{\square} + 2) \psi,     
\label{Dirac-squared}
\end{eqnarray}
defining the covariant 4-dimensional Laplacian $\widehat{\square}$ 
acting on the spinors. This corresponds to the usual expression 
for $\widehat{D}^2$ on curved spaces, and  
the constant 2 is due to the curvature scalar. 
Since $\widehat{D}^2$ and $\Sigma^{\mu\nu}F_{\mu\nu}$ are both Hermitian
and commute with $\G$ and $\widehat{P}_{\pm}$ 
as defined in \eq{projection operator 1} in the large $N$ limit, 
it follows that $\widehat{\square}$ satisfies these properties as well. 
Note that \eq{Dirac-squared} can also be
written as
\be
(\widehat{D} - C_0)^2 =  \Sigma^{\mu\nu}F_{\mu\nu}+
\widehat{\mathcal{D}}_{\mu}\widehat{\mathcal{D}}_{\mu} + \frac 12,
\ee
which might suggest to interpret
$\widehat{\mathcal{D}}_{\mu}\widehat{\mathcal{D}}_{\mu}$ as covariant 
Laplacian; however this is not correct since 
$\widehat{\mathcal{D}}_{\mu}\widehat{\mathcal{D}}_{\mu}$ does not
commute with the projections $\widehat{P}_{\pm }$ \eq{projection operator 1}
even in the commutative limit. The
reason for this is our formulation using spinors  
based on the $SO(6)$ Clifford algebra
rather than $SO(4)$ spinors and comoving frames. The corresponding
projections to physical Dirac- or Weyl-spinors in the fuzzy case will
be discussed next.

\subsubsection{Projections for the fuzzy spinors}
\label{chapter on fuzzy projections}

For the fuzzy case, we can again consider the following projection operators
\begin{eqnarray*}
\widehat{\chi }_{L}\Psi  & = & \frac{2}{N}(\Gamma ^{iL}\Psi \lambda _{iL}+C_0^L\Psi ),\\
\widehat{\chi }_{R}\Psi  & = & \frac{2}{N}(\Gamma ^{iR}\Psi \lambda _{iR}+C_0^R\Psi )
\end{eqnarray*}
which satisfy
\[
\widehat{\chi }_{L,R}^{2}=1, 
\qquad \{\widehat{\chi }_{L},\widehat{\chi }_{R}\}=0\, .
\]
This implies 
$(\widehat{\chi }_{L}\widehat{\chi }_{R})^{2}=-1$,
and we can write down the following projection operators  
\begin{equation}
\label{projection operator 1}
\widehat{P}_{\pm }=\frac{1}{2}(1\pm i\widehat{\chi }_{L}\widehat{\chi }_{R})
\end{equation}
which have the classical limit (\ref{projector})
and the properties (\ref{properties of p+-}). 
However, the projector no longer commutes with the fuzzy Dirac operator
(\ref{fuzzy dirac operator}):
\begin{eqnarray*}
{}[\widehat{D},\widehat{\chi }_{L}\widehat{\chi }_{R}] & = & 
\{\widehat{D},\widehat{\chi }_{L}\}\widehat{\chi }_{R}-\widehat{\chi }_{L}\{\widehat{D},\widehat{\chi }_{R}\}\\
 & = & -\frac{2}{N}\Big(\big (2(\lambda _{i L}+\mc
 A_{iL})\widehat{J}_{i L}-2 A_{i L}\lambda _{iL}
+ 2 C_0^L\; \Gamma ^{iL}\widehat{\mathcal{D}}_{iL}+1\big )\widehat{\chi }_{R}\\
 &  & -\widehat{\chi }_{L}\big (2(\lambda _{iR}+\mc
 A_{iR})\widehat{J}_{iR}-2 A_{iR}\lambda _{iR}
+ 2 C_0^R\;\Gamma ^{iR}\widehat{\mathcal{D}}_{iR}+1\big )\Big)\, ,
\end{eqnarray*}
 which only vanishes for \( N\rightarrow \infty  \) and 
tangential \( A_{\mu } \) \eq{A-constraint}.
To reduce the degrees of freedom to one Dirac 4-spinor, 
we should therefore add
a mass term 
\be
M_- \Psi_{8}^{\dagger }\widehat{P}_{-}\Psi _{8}
\label{M-constraint-fuzzy}
\ee
which for \( M_- \rightarrow \infty  \) suppresses one of the spinors,
rather than impose an exact constraint as in (\ref{constraint}). 
This is  gauge invariant 
since $\widehat{P}_{\pm }$ commutes with gauge transformations, 
\[
\widehat{P}_{\pm }\psi \rightarrow U\widehat{P}_{\pm }\psi \, .
\]
The complete action for a Dirac fermion 
on fuzzy $S^2_N \times S^2_N$
is therefore given by 
\be
S_{\mathrm{Dirac}} = \int \Psi _{8}^{\dagger }(\widehat{D} + m)\Psi _{8} + 
M_- \Psi _{8}^{\dagger }\widehat{P}_{-}\Psi _{8}
\ee
with $M_- \to \infty$.
The physical chirality operator is given by 
$\G$ \eq{chiral-G-op},
which allows to consider Weyl spinors as well.

\section{Conclusion and outlook} 

We have constructed  $U(n)$ gauge theory on fuzzy $S^2_N \times
S^2_N$ as a multi-matrix model. The model is completely finite, and can be
considered as a regularization either of Yang-Mills 
on the commutative $S^2 \times
S^2$, or on the noncommutative $\R^4_\theta$ in a suitable scaling
limit.  The quantization is defined by a finite ``path'' integral
over the matrix degrees of freedom, which is  convergent
due to the constraint term. A
gauge-fixed action with BRST symmetry is also provided.
We then discussed some topologically non-trivial solutions
in the $U(1)$ case, which reduce to the known ``fluxon''
solutions on $\R^4_\theta$ in the appropriate scaling limit,
reproducing the full moduli space. 
On $S^2_N\times S^2_N$
they arise as localized flux tubes together with a monopole background field.
This provides a very clean non-perturbative definition of 
noncommutative gauge theory with fixed rank of the gauge group $U(n)$,
and a simple description of instantons as
solutions of the equation of motion in one single configuration space.
Furthermore, we have shown how charged fermions 
in the fundamental representation can be coupled to the
gauge field, by defining a suitable Dirac operator
$\widehat{D}$. This is easily extended to 
Weyl fermions using a chirality operator which exactly anticommutes 
with $\widehat{D}$.
All this supports the programme to formulate and study
physically interesting 
models on noncommutative spaces.

There are many interesting conclusions and applications to be
explored.
One crucial feature is the fact that the model is completely 
regularized, i.e. the quantization is well-defined
without any divergences for finite $N$. This should allow to study 
suitable scaling limits in $N$ in a well-defined framework, 
and the emergence of an interesting low-energy limit which could be
either commutative or noncommutative. Such a matrix regularization 
is very interesting in view
of the UV/IR mixing, which indicates a close relationship 
between NC field theory and matrix models. 
For example, one might try to extend the results
in \cite{Steinacker:2005wj} in this context. 
We also explored some alternative formulations 
using ``collective matrices'' based on $SO(6)$. 
Such formulations are possible only in the noncommutative case, and 
lead to the hope that new non-perturbative techniques 
in the spirit of random matrix theory
may be developed along these lines. 

Another important aspect is the
coupling to fermions, which could be extended to scalars and
allows to study spontaneous symmetry breaking
and the possible generation of other gauge groups in the low-energy 
limit. Finally, a detailed comparison with other finite models of NC gauge
theory in 4 dimensions 
such as \cite{Grosse:2004wm,Ambjorn:2000nb,Ambjorn:2000cs}
would be very desirable, to see which features are generic and which
are model-dependent.

\subsection*{Acknowledgements} 

We would like to thank Andreas Sykora and Badis Ydri for collaboration
and discussions on the 2-dimensional case, in an early stage of this
project. We also want to thank Julius Wess for his steady support.

\appendix

\section{The standard representation of the fuzzy sphere}
\label{sec:useful}

The irreducible  $N$-dimensional representation of the 
$su(2)$ algebra $\lambda_i$ \eq{lambda-algebra} is given by
\bea \label{reps}
&&(\lambda_3)_{kl} = \d_{kl} \; \frac{N+1-2k}{2},\\
&&(\lambda_+)_{kl} = \d_{k+1,l}\sqrt{(N-k)k},
\label{standard-rep}
\eea
where $k,l = 1, ..., N$ and  $\lambda_\pm = \lambda_1 \pm i\lambda_2$.

\section{Alternative formulation using $4\cN \times 4\cN$ matrices}
\label{sec:4Nmatrices}

Let us rewrite the action \eq{eq: action S_6 - S_break} in terms of the 
$4\cN \times 4\cN$ matrices $B_L, B_R$  \eq{B-L-R}.
Noting that
\be
C_L C_R + C_R C_L
=  \left(\begin{array}{cc} -[B_L, B_R] & 0 \nn\\ 
 0 & [B_L, B_R]
  \end{array}\right)
\ee
we can rewrite $S_6$ \eq{action-so6} as 
\be
S_6 = 2 \mathrm{Tr}
\left(B_L^2 - B_R^2 -\frac{N^2}2\right)^2 
+ 2 \mathrm{Tr}\left([B_L,B_R]^2\right),
\ee
where the trace is now over $4\cN \times 4\cN$ matrices.
Similarly
\bea 
S_{\mathrm{break}} &=& - 2 \mathrm{Tr}\left(B_L^2 - \frac{N^2}4\right) \left(-B_R^2
  - \frac{N^2}4\right) 
\eea 
and combined we recover \eq{action} as
\be 
S =  S_{6}-2 S_{\mathrm{break}}  
= 2 \mathrm{Tr} \left((B_L^2 -\frac{N^2}4)^2
  + (-B_R^2 -\frac{N^2}4)^2 + [B_L,B_R]^2 \right).  
\label{action-BLR}
\ee   
This looks like a 2-matrix model, however the degrees of freedom 
$B_L, B_R$ are still very much constrained and  span 
only a small subspace of the  
$4\cN \times 4 \cN $ matrices. We
would like to find an intrinsic characterization without using the 
$\g_\mu$ explicitly. One possibility is to choose the $\g_\mu$ to be
completely anti-symmetric matrices, see Appendix \ref{app:gamma-mat}.
However this does not extend to $B$,
since the $B_\mu$ should be Hermitian and not necessarily symmetric, and
moreover the $\g_\mu$ are not Hermitian (the conjugate being the
intertwiner $(6) \subset (\obar 4) \tens (\obar 4)$). Another
possibility is provided by the following representation of the
$\g$-matrices: 
\be
\g_L^i = \sigma^i \tens \one_{2\times 2}, \qquad
\g_R^i =  \one_{2\times 2} \tens i \sigma^i.
\label{gamma-rep-1}
\ee
They satisfy the relations \eq{g-conj} -- \eq{g-L-R}, 
but are not antisymmetric.
Now note that 
\be
\g_R^i = i P \g_L^i P
\ee
where
\beq 
P = \left(\begin{array}{cccc} 1 & 0 & 0 & 0 \\ 0 & 0 & 1 & 0 
\\ 0 & 1 & 0 & 0 \\ 0 & 0 & 0 & 1 \end{array} \right) 
= \frac 12 (1+\sigma^i \tens \sigma^i) 
\label{P-explicit} 
\eeq 
permutes the two tensor factors and satisfies 
\beq 
P^2 = 1. 
\eeq 
Therefore we can characterize the degrees of freedom
in terms of 2 Hermitian $2\cN \times 2\cN$ matrices 
\be
X_L = B_L^i \sigma_i + \frac 12, \qquad
X_R = B_R^i \sigma_i + \frac 12 
\ee
which are arbitrary up to the constraint that 
$X_{L,R}^0 = \frac 12$. Then 
\be
B_L = X_L \tens \one_{2\times 2}, \quad
B_R = i P (X_R \tens \one_{2\times 2}) P;
\ee
they could be extracted from  a single complex matrix
$\tilde B =  (X_L+i X_R) \tens \one_{2\times 2}$.
Furthermore, matrices of the form $X \tens \one_{2\times 2}$
are characterized through their spectrum, 
which is doubly degenerate;
indeed any such Hermitian matrix can be cast into the above form
using suitable unitary $SU(4\cN)$ transformations. 
Similarly, $P$ can also be characterized intrinsically:
any matrix $P$ written as 
\be 
P = P_0 \tens \one_{2\times 2} + P_i \tens \sigma^i 
\ee 
which satisfies the constraints 
\be P_0 = \frac 12, \quad P^2 = \one
\ee 
is given by \eq{P-explicit} up to an irrelevant unitary transformation
$U \tens \one$.
We could therefore write down the action \eq{action-BLR} 
in terms of three matrices $B_L, -iP B_R P$ and $P$, all of
which are characterized by
their spectrum and constraints of the form $(..)_0 = \frac 12$.
The hope is that such a reformulation may allow to
apply some of the powerful
methods from random matrix theory, in the spirit of \cite{Steinacker:2003sd}. 
However we will leave this for
future investigations.

\section{Stability analysis of the $SO(6)$ - invariant action \eq{action-so6}}\label{app:stability}

Consider the action \eq{action-so6}. 
We will split off the
radial degrees of freedom for large $N$ by  setting $R=1$ and\footnote{The fact
  that this leads to non-hermitian fields for finite $N$ is not
essential here.}
\[
B_{iL}= \lambda _{iL}+ A_{iL} =  \lambda _{iL}+ \mc A_{iL}+x_{iL}\Phi _{L}\]
requiring that $\lambda_{iL} \mc A_{iL} =0$,
and similarly for $B_{iR}$,
The stability of our geometry will depend on the behavior of \(\Phi^L\) and \(\Phi^R\). We calculate that

\[
B_{\mu }B_{\mu }-\frac{N^{2}-1}{2}=N(\Phi _{L}+\Phi _{R})+\Phi _{L}\Phi _{L}+\Phi _{R}\Phi _{R}+\mc A_{\mu }\mc A_{\mu }-[\lambda _{\mu },\mc A_{\mu }]+ O(\frac{1}{N}),\]
 where we used that \( \lambda _{ia}\mc A_{ia}=0 \) and therefore
both \( \mc A_{ia}x_{ia}= O(\frac{1}{N}) \) and 
\( \mc A_{ia}[\lambda _{ia},\: \cdot \: ]= O(\frac{1}{N}) \) for $a=L,R$.
Setting 
\begin{eqnarray*}
\Phi _{L}+\Phi _{R} & = & \Phi _{1},\\
\Phi _{L}-\Phi _{R} & = & \Phi _{2}
\end{eqnarray*}
we get
\begin{equation}
\label{B_i B_i}
B_{\mu }B_{\mu }-\frac{N^{2}-1}{2}=N\Phi _{1}+\Phi _{1}\Phi _{1}+\Phi _{2}\Phi _{2}+\mc A_{\mu }\mc A_{\mu }-[\lambda _{\mu },\mc A_{\mu }]+ O(\frac{1}{N}).
\end{equation}
 In the limit \( N\rightarrow \infty  \) we can integrate out
\( \Phi _{1} \), as it acquires an infinite mass. Alternatively we
can rescale \( \Phi _{1} \) by setting \( \phi _{1}=\frac{1}{N}\Phi _{1} \).
Then, all the terms involving \( \phi _{1} \) but the first one in
(\ref{B_i B_i}) will be of order \( \frac{1}{N} \) and we can equally
integrate out \( \phi _{1} \). 

The terms from
\[
F_{\mu }F_{\mu }-[B_{iL},B_{iR}]^{2}
\]
involving the remaining \( \Phi _{2} \) will be (in the limit \(
N\rightarrow \infty  \))
\[
\frac{1}{2}\Phi _{2}\Phi _{2}-J_{\mu }(\Phi _{2})J_{\mu }(\Phi
_{2})-F_{iL}x_{iL}\Phi _{2}+F_{iR}x_{iR}\Phi _{2}
\]
with the tangential derivatives \( J_{ia}=-i\epsilon _{ijk}x_{ja}\dd _{ka} \).
Calculating that
\[
J_{\mu }\Phi _{2}J_{\mu }\Phi _{2}=-\dd _{\mu }\Phi _{2}\dd _{\mu
}\Phi _{2}-x_{iL}\dd _{iL}\Phi _{2}x_{jL}\dd _{jL}\Phi _{2}-x_{iR}\dd
_{iR}\Phi _{2}x_{jR}\dd _{jR}\Phi _{2}
\]
and using partial integration under the integral this gives
\[
\frac{1}{2}\Phi _{2}\Phi _{2}-\Phi _{2}\dd _{\mu }\dd _{\mu }\Phi
_{2}-x_{iL}\dd _{iL}\Phi _{2}x_{jL}\dd _{jL}\Phi _{2}-x_{iR}\dd
_{iR}\Phi _{2}x_{jR}\dd _{jR}\Phi _{2}-F_{iL}x_{iL}\Phi
_{2}+F_{iR}x_{iR}\Phi _{2}
\]
Expanding both \( \Phi _{2} \) and \( F \) in left and right spherical
harmonics as
\[
\Phi _{2}=\sum _{klmn}c_{klmn}Y^{L}_{km}Y^{R}_{ln}\; \; \; \; \mbox
{and}\; \; \; \; F_{ia}x_{ia}=\sum
_{klmn}f^{a}_{klmn}Y^{L}_{km}Y^{R}_{ln}
\]
 we get for fixed \( klmn \), setting \( c=c_{klmn} \), \( f^{a}=f^{a}_{klmn} \)
and \( p=\frac{1}{2}+l(l+1)+k(k+1) \) the following expression
\[
pc^{2}-cf^{L}+cf^{R}=p(c-\frac{1}{2p}f^{L}+\frac{1}{2p}f^{R})^{2}-\frac{1}{4p}(f^{L}-f^{R})^{2}.
\]
 Integrating out the \( c \)'s and putting everything back this leaves
us with the additional term
\[
-(F_{iL}x_{iL}-F_{iR}x_{iR})\frac{1}{4(\frac{1}{2}-\dd_{\mu }\dd
  _{\mu })}(F_{iL}x_{iL}-F_{iR}x_{iR})
\]
in the action \eq{action-so6}.

\section{Representation of the \protect\( SO(6)\protect \)- intertwiners
  and Clifford algebra}\label{app:gamma-mat}

We will use the Pauli matrices
\[
\sigma ^{1}=\left( \begin{array}{cc}
0 & 1\\
1 & 0
\end{array}\right) ,\hspace {0,8cm}\sigma ^{2}=\left( \begin{array}{cc}
0 & -i\\
i & 0
\end{array}\right) ,\hspace {0,8cm}\sigma ^{3}=\left( \begin{array}{cc}
1 & 0\\
0 & -1
\end{array}\right) ,\]
which satisfy
\be
\sigma^i \sigma^j = \d^{ij} + i \vare^{ijk} \sigma^k.
\ee
 With these we define the \( 4 \)-dimensional antisymmetric matrices
\be \barr{llll}
&\gamma ^{1}_{L}=\sigma ^{1}\otimes \sigma ^{2},
&\gamma^{2}_{L}=\sigma ^{2}\otimes 1,
&\gamma ^{3}_{L}=\sigma^{3}\otimes \sigma ^{2}, \\
&\gamma ^{1}_{R}=i\; \sigma ^{2}\otimes \sigma ^{1},
&\gamma ^{2}_{R}=i\; 1\otimes \sigma ^{2},
&\gamma^{3}_{R}=i\; \sigma ^{2}\otimes \sigma ^{3}.
\earr
\label{gamma-wolfg}
\ee
They are the intertwiners between \( SU(4)\otimes SU(4) \) and \( SO(6) \)
and fulfill the following relations:
\begin{eqnarray*}
(\gamma ^{i}_{L})^{\dagger } & = & \gamma ^{i}_{L},\\
(\gamma ^{i}_{R})^{\dagger } & = & -\gamma ^{i}_{R}
\end{eqnarray*}
and
\begin{eqnarray*}
\gamma _{L}^{i}\gamma _{L}^{j} & = & \delta ^{ij}+i\epsilon ^{ij}_{k}\gamma _{L}^{k},\\
\gamma _{R}^{i}\gamma ^{j}_{R} & = & -\delta ^{ij}-\epsilon ^{ij}_{k}\gamma ^{k}_{R},\\
{}[\gamma ^{i}_{L},\gamma ^{j}_{R}] & = & 0.
\end{eqnarray*}
We can now define the \( 8 \)-dimensional representation of the 
\( SO(6) \)-Clifford
algebra as
\begin{equation}
%\label{Gamma}
\Gamma ^{\mu }=\left( \begin{array}{cc}
0 & \gamma ^{\mu }\\
\gamma ^{\mu \dagger } & 0
\end{array}\right) ,
\end{equation}
with the desired anticommutation relations 
\[
\{\Gamma ^{\mu },\Gamma ^{\nu }\}=\left( \begin{array}{cc}
\gamma ^{\mu }\gamma ^{\nu \dagger }+\gamma ^{\nu }\gamma ^{\mu \dagger } & 0\\
0 & \gamma ^{\mu \dagger }\gamma ^{\nu }+\gamma ^{\nu \dagger }\gamma ^{\mu }
\end{array}\right) =2\delta ^{\mu \nu }.
\]
 The chirality operator in this basis is
\[
\Gamma =i\Gamma ^{1}_{L}\Gamma ^{2}_{L}\Gamma ^{3}_{L}\Gamma ^{1}_{R}\Gamma _{R}^{2}\Gamma ^{3}_{R}=\left( \begin{array}{cc}
-1 & 0\\
0 & 1
\end{array}\right) .
\]
The \( 8 \)-dimensional \( SO(6) \)-rotations are generated by\[
\Sigma _{8}^{\mu \nu }=-\frac{i}{4}[\Gamma ^{\mu },\Gamma ^{\nu }]=-\frac{i}{4}\left( \begin{array}{cc}
\gamma ^{\mu }\gamma ^{\nu \dagger }-\gamma ^{\nu }\gamma ^{\mu \dagger } & 0\\
0 & \gamma ^{\mu \dagger }\gamma ^{\nu }-\gamma ^{\nu \dagger }\gamma ^{\mu }
\end{array}\right) .
\]

\section{The\label{appendix b} Dirac operator in spherical coordinates}

For a general Riemannian manifold with metric\[
g=g_{\mu \nu }dx^{\mu }dx^{\nu }\]
 the Christoffel symbols are given by\begin{equation}
\label{Christoffel symbol}
\Gamma ^{\sigma }_{\mu \nu }=\frac{1}{2}g^{\sigma \lambda }(\pp \mu g_{\lambda \nu }+\pp \nu g_{\lambda \mu }-\pp \lambda g_{\mu \nu }).
\end{equation}
We can change to a non-coordinate basis (labeled by latin indices
in contrast to the greek indices for the coordinates) by introducing
the vielbeins \( e^{\mu }_{a} \) with 
\begin{eqnarray*}
e^{a}_{\mu }e^{\mu }_{b} & = & \delta ^{a}_{b},\\
g_{\mu \nu } & = & e^{a}_{\mu }e_{\nu }^{b}\delta _{ab}, \qquad
g^{\mu \nu }  =  e^{\mu }_{a}e^{\nu }_{b}\delta ^{ab}.
\end{eqnarray*}
With these, the Dirac operator is given by\[
D=-i\gamma ^{a}e_{a}^{\mu }(\pp \mu +\frac{1}{4}\omega _{\mu ab}[\gamma ^{a},\gamma ^{b}]),\]
where the \( \gamma ^{a} \) form a flat Clifford algebra, i. e.\[
\{\gamma ^{a},\gamma ^{b}\}=2\delta ^{ab}\; \; \; \; ,\; \; \; \; \gamma ^{a\dagger }=\gamma ^{a}\]
 and the spin connection \( \omega  \) fulfills \begin{equation}
\label{spinor connection}
\pp \mu e^{a}_{\nu }-\Gamma ^{\lambda }_{\mu \nu }e_{\lambda }^{a}+\omega _{\mu \: \: b}^{\: \: a}\: e^{b}_{\nu }=0.
\end{equation}

\subsection{The Dirac operator on \protect\( \mathbb R^{6}\protect \)
in spherical coordinates}

We will now write down the flat \( SO(6) \) Dirac operator \( D_{6} \)
by splitting \( \mathbb R^{6} \) into \( \mathbb R_{L}^{3}\times \mathbb R_{R}^{3} \)
and introducing spherical coordinates on both the left and right hand
side. The flat metric becomes\begin{eqnarray}
g_{6} & = & r_{L}^{2}\; d\theta _{L}\otimes d\theta _{L}+r_{L}^{2}\sin ^{2}\theta _{L}\; d\phi _{L}\otimes d\phi _{L}+dr_{L}\otimes dr_{L}\label{metric 6D} \\
 &  & +r_{R}^{2}\; d\theta _{R}\otimes d\theta _{R}+r_{R}^{2}\sin ^{2}\theta _{R}\; d\phi _{R}\otimes d\phi _{R}+dr_{R}\otimes dr_{R}.
\end{eqnarray}
 Looking at the formula for the Christoffel symbols (\ref{Christoffel symbol}),
we see that all the symbols with both right and left indices vanish.
For the symbols with only right or only left indices we get\begin{eqnarray}
\Gamma ^{\theta }_{\phi \phi } & = & -\sin \theta \cos \theta ,\label{christoffel 6D} \\
\Gamma ^{\phi }_{\theta \phi } & = & \frac{\cos \theta }{\sin \theta }=\Gamma ^{\phi }_{\phi \theta ,}\label{christoffel 6D 2} \\
\Gamma ^{r}_{\theta \theta } & = & -r,\\
\Gamma ^{r}_{\phi \phi } & = & -r\sin ^{2}\theta ,\\
\Gamma ^{\theta }_{r\theta } & = & \frac{1}{r}=\Gamma ^{\theta }_{\theta r},\\
\Gamma ^{\phi }_{r\phi } & = & \frac{1}{r}=\Gamma ^{\phi }_{\phi r},
\end{eqnarray}
where we have dropped the left or right subscript for simplicity.
All other symbols vanish. We want to go to a non-coordinate basis
by introducing the vielbeins\begin{eqnarray}
e_{\theta _{L}}^{1_{L}}=r_{L}; & e^{2_{L}}_{\phi _{L}}=r_{L}\sin \theta _{L}; & e^{3_{L}}_{r_{L}}=1;\label{vielbein 6D} \\
e_{\theta _{R}}^{1_{R}}=r_{L}; & e^{2_{R}}_{\phi _{R}}=r_{R}\sin \theta _{R}; & e^{3_{R}}_{r_{R}}=1.\label{vielbein 6D 2} 
\end{eqnarray}
Calculating the spinor connection by (\ref{spinor connection}), we
again see that all the terms with both left and right indices vanish.
The terms with only left or only right indices are \begin{eqnarray}
\omega _{\phi \: \: 2}^{\: \: 1}= & -\cos \theta  & =-\omega _{\phi \: \: 1}^{\: \: 2},\label{spin conn 6D} \\
\omega ^{\: \: 2}_{\phi \: \: 3}= & \sin \theta  & =-\omega ^{\: \: 3}_{\phi \: \: 2},\\
\omega _{\theta \: \: 3}^{\: \: 1}= & 1 & =-\omega _{\theta \: \: 1}^{\: \: 3},
\end{eqnarray}
where we again dropped the left or right subscripts. Putting all this
together we see that \( D_{6} \) splits up into a left part \( D_{3L} \)
and a right part \( D_{3R} \) as 
\be
D_{6}=D_{3L}+D_{3R}
\label{D6-3plus3}
\ee
with 
\begin{eqnarray}
D_{3L} & = & -i\overline{\Gamma }_{L}^{1}\frac{1}{r_{L}}(\dd _{\theta _{L}}+\frac{\cos \theta _{L}}{\sin \theta _{L}})-i\overline{\Gamma }^{2}_{L}\frac{1}{r_{L}\sin \theta _{L}}\dd _{\phi _{L}}-i\overline{\Gamma }^{3}_{L}(\dd _{r_{L}}+\frac{1}{r_{L}}),\label{D3L spherical} \\
D_{3R} & = & -i\overline{\Gamma }_{R}^{1}\frac{1}{r_{R}}(\dd _{\theta _{R}}+\frac{\cos \theta _{R}}{\sin \theta _{R}})-i\overline{\Gamma }^{2}_{R}\frac{1}{r_{R}\sin \theta _{R}}\dd _{\phi _{R}}-i\overline{\Gamma }^{3}_{R}(\dd _{r_{R}}+\frac{1}{r_{R}}).\label{D3R spherical} 
\end{eqnarray}
where the \( \overline{\Gamma } \) have to form a \( SO(6) \) Clifford
algebra.

\subsection{The Dirac operator on \protect\( S^{2}\times S^{2}\protect \) }

We now want to calculate the curved Dirac operator \( D_{4} \) on
\( S^{2}\times S^{2} \) in the spherical coordinates of the spheres
(they are the same spherical coordinates we used before, now restricted
to the spheres). The metric on \( S^{2}\times S^{2} \) with radii
\( r _{L} \) and \( r _{R} \) is\begin{eqnarray*}
g_{4} & = & r _{L}^{2}\; d\theta _{L}\otimes d\theta _{L}+r _{L}^{2}\sin ^{2}\theta _{L}\; d\phi _{L}\otimes d\phi _{L}\\
 &  & +r _{R}^{2}\; d\theta _{R}\otimes d\theta _{R}+r _{R}^{2}\sin ^{2}\theta _{R}\; d\phi _{R}\otimes d\phi _{R}.
\end{eqnarray*}
The metric is the same as (\ref{metric 6D}) restricted to the spheres,
so the Christoffel symbols are the same as (\ref{christoffel 6D})
and (\ref{christoffel 6D 2}). Again introducing the vielbeins\begin{eqnarray}
e_{\theta _{L}}^{1_{L}}=r _{L}; & e^{2_{L}}_{\phi _{L}}=r _{L}\sin \theta _{L}; & \label{vielbein 4D} \\
e_{\theta _{R}}^{1_{R}}=r _{L}; & e^{2_{R}}_{\phi _{R}}=r _{R}\sin \theta _{R}, & 
\end{eqnarray}
we see that also the spin connection is the same as (\ref{spin conn 6D}),
and therefore we can again split \( D_{4} \) into a right part \( D_{2R} \)
and a left part \( D_{2L} \) as \( D_{4}=D_{2L}+D_{2R} \) with

\begin{eqnarray}
D_{2L} & = & -i\widetilde{\Gamma }_{L}^{1}\frac{1}{r _{L}}(\dd _{\theta _{L}}+\frac{\cos \theta _{L}}{\sin \theta _{L}})-i\widetilde{\Gamma }^{2}_{L}\frac{1}{r _{L}\sin \theta _{L}}\dd _{\phi _{L}},\label{D2-L}\\
D_{2R} & = & -i\widetilde{\Gamma }_{R}^{1}\frac{1}{r _{R}}(\dd
_{\theta _{R}}+\frac{\cos \theta _{R}}{\sin \theta_{R}})
-i\widetilde{\Gamma }^{2}_{R}
\frac{1}{r _{R}\sin \theta_{R}}\dd _{\phi _{R},}
\label{D2-R}
\end{eqnarray}
where the \( \widetilde{\Gamma } \) form a flat \( SO(4) \) Clifford
algebra.

\subsection{$SO(3)\times SO(3)$-covariant form of the
Dirac operator on \protect\( S^{2}\times S^{2}\protect \) }
\label{sec:so6-dirac-4}

The flat \( SO(6) \) Dirac operator \( D_{6} \) was split
into a left part \( D_{3L} \) and
a right part \( D_{3R} \) using spherical coordinates in \eq{D6-3plus3}.
Of course, \( D_{6} \)
can also be written in the usual Euclidean coordinates as
\[
D_{6}=-i\Gamma ^{\mu }\pp \mu ,
\]
 where again we can split it into a left and a right part as 
\[
D_{6}=D_{3L}+D_{3R}
\]
 with\begin{eqnarray}
&&D_{3L}  =  -i\Gamma ^{i}_{L}\pp i, \qquad
D_{3R}  =  -i\Gamma _{R}^{i}\pp i,\label{D3 euklid} \\
&& \qquad \qquad \{D_{3L},D_{3R}\}  =  0.
\end{eqnarray}
We have left open which representation of the \( SO(6) \) Clifford
algebra we want to use for the \( \overline{\Gamma } \) 
in (\ref{D3L spherical},\ref{D3R spherical}),
but \( \Gamma  \) in (\ref{D3 euklid}) is really
the representation given by (\ref{Gamma}). We will now relate the
two expressions for the Clifford algebra and the Dirac operator by
first defining 
\[
J_{iL}=-i\epsilon _{ijk}x_{jL}\dd _{kL}\; \; \; \; \mbox {and}\; \; \;
\; J_{iR}=-i\epsilon _{ijk}x_{jR}\dd _{kR}
\]
 and noting that
\begin{equation}
\label{eins calc}
\left( \frac{\Gamma ^{i}_{L}x_{iL}}{r_{L}}\right) ^{2}=\left( \frac{\Gamma ^{i}_{R}x_{iR}}{r_{R}}\right) ^{2}=1.
\end{equation}
 We calculate that 
\begin{eqnarray}
\left( \frac{\Gamma^{j}_{L}x_{jL}}{r_{L}}\right)^{2}\Gamma^{i}_{L}\dd _{iL} & = & \left( \frac{\Gamma^{j}_{L}x_{jL}}{r_{L}}\right) \left( \frac{x_{iL}\dd _{iL}}{r_{L}}-\frac{1}{r_{L}}\left( \begin{array}{cc}
\gamma ^{i}_{L} & 0\\
0 & \gamma ^{i}_{L}
\end{array}\right) J_{iL}\right) ,\label{calc gamma L} \\
\left( \frac{\Gamma^{j}_{R}x_{jR}}{r_{R}}\right) ^{2}\Gamma^{i}_{R}\dd _{iR} & = & \left( \frac{\Gamma^{j}_{R}x_{jR}}{r_{R}}\right) \left( \frac{x_{iR}\dd _{iR}}{r_{R}}+\frac{i}{r_{R}}\left( \begin{array}{cc}
\gamma ^{i}_{R} & 0\\
0 & \gamma ^{i}_{R}
\end{array}\right) J_{iR}\right) ,\label{calc gamma R} 
\end{eqnarray}
and therefore 
\begin{eqnarray}
D_{3L} & = & -i\left( \frac{\Gamma^{j}_{L}x_{jL}}{r_{L}}\right) \left( \dd _{r_{L}}-\frac{1}{r_{L}}\left( \begin{array}{cc}
\gamma ^{i}_{L} & 0\\
0 & \gamma ^{i}_{L}
\end{array}\right) J_{iL}\right) ,\label{calc D3L} \\
D_{3R} & = & -i\left( \frac{\Gamma^{j}_{R}x_{jR}}{r_{R}}\right) \left( \dd _{r_{R}}+\frac{i}{r_{R}}\left( \begin{array}{cc}
\gamma ^{i}_{R} & 0\\
0 & \gamma ^{i}_{R}
\end{array}\right) J_{iR}\right) .\label{calc D3R} 
\end{eqnarray}
Comparing this with (\ref{D3L spherical},\ref{D3R spherical}) we
see that 
\begin{equation}
\label{Gamma 3 calc}
\overline{\Gamma }^{3}_{L}=\left( \frac{\Gamma ^{i}_{L}x_{iL}}{r_{L}}\right) \; \; \; \; \mbox {and}\; \; \; \; \overline{\Gamma }^{3}_{R}=\left( \frac{\Gamma ^{i}_{R}x_{iR}}{r_{R}}\right) ,
\end{equation}
as the \( J_{L} \) and \( J_{R} \) have no radial components. From
(\ref{calc D3L},\ref{calc D3R}) we can also deduce that
\begin{equation}
\label{comm}
[\overline{\Gamma }^{i}_{L},\left( \begin{array}{cc}
0 & 1\\
1 & 0
\end{array}\right) ]=0=[\overline{\Gamma }^{i}_{R},\left( \begin{array}{cc}
0 & 1\\
-1 & 0
\end{array}\right) ]
\end{equation}
and\begin{equation}
\label{anti}
\{\overline{\Gamma }^{i}_{R},\left( \begin{array}{cc}
0 & 1\\
1 & 0
\end{array}\right) \}=0=\{\overline{\Gamma }^{i}_{L},\left( \begin{array}{cc}
0 & 1\\
-1 & 0
\end{array}\right) \}.
\end{equation}
 The curved Dirac operator \( D_{4} \) on \( S^{2}\times S^{2} \)
expressed in the spherical coordinates of the spheres also splits
up  as
\( D_{4}=D_{2L}+D_{2R} \) with right part \( D_{2R} \) and 
 left part \( D_{2L} \) given in \eq{D2-L},\eq{D2-R}.
Comparing this with (\ref{D3L spherical},\ref{D3R spherical}),
we see that the dependence on the tangential coordinates is the same
in both expressions. With (\ref{comm},\ref{anti}) we see that the
matrices \( -i\left( \begin{array}{cc}
0 & 1\\
1 & 0
\end{array}\right) \overline{\Gamma }^{3}_{L}\overline{\Gamma }^{i}_{L} \) and \( \left( \begin{array}{cc}
0 & 1\\
-1 & 0
\end{array}\right) \overline{\Gamma }^{3}_{R}\overline{\Gamma }^{j}_{R} \) for \( i,j=1,2 \) form a \( SO(4) \) Clifford algebra and can therefore
be used as the \( \widetilde{\Gamma } \). Note that this representation
is still reducible, a problem we deal with in Section
\ref{chapter on projections}.
Now we can get a simple relation between the \( D_{3} \) restricted
on the spheres and the \( D_{2} \)
\begin{eqnarray*}
-\left( \begin{array}{cc}
0 & 1\\
1 & 0
\end{array}\right) (i\overline{\Gamma }^{3}_{L}D_{3L}|_{res.}-\frac{1}{r _{L}}) & = & D_{2L},\\
-i\left( \begin{array}{cc}
0 & 1\\
-1 & 0
\end{array}\right) (i\overline{\Gamma }^{3}_{R}D_{3R}|_{res.}-\frac{1}{r _{R}}) & = & D_{2R}.
\end{eqnarray*}
 Inserting (\ref{calc D3L},\ref{calc D3R}) and using (\ref{Gamma 3 calc})
together with (\ref{eins calc}) we find that
\begin{eqnarray}
D_{2L} & = & \frac{1}{r _{L}}(\Gamma ^{i}_{L}J_{iL}+\left( \begin{array}{cc}
0 & 1\\
1 & 0
\end{array}\right) ),\label{D 2L final} \\
D_{2R} & = & \frac{1}{r _{R}}(\Gamma ^{i}_{R}J_{iR}+i\left( \begin{array}{cc}
0 & 1\\
-1 & 0
\end{array}\right) ).
\label{D 2R final} 
\end{eqnarray}
Setting \( r _{L}=r _{R}=1 \) for simplicity,
the Dirac operator \( D_4 \) 
on \( S^{2}\times S^{2} \) takes the form \eq{Dirac operator}.

\bibliographystyle{diss}

\bibliography{mainbib}

\end{document}